\documentclass[manuscript]{acmart}
\usepackage{graphicx}
\usepackage{subcaption}
\AtBeginDocument{%
  }

\setcopyright{acmlicensed}
\copyrightyear{2025}
\acmYear{2025}
\acmDOI{}

\acmISBN{978-1-4503-XXXX-X/18/06}

\begin{document}

\title{"They Aren't Built For Me": A Replication Study of Visual Graphical Perception with Tactile Representations of Data for Visually Impaired Users}

%
\author{Areen Khalaila}
\email{areenkh@brandeis.edu}
\affiliation{%
  \institution{Brandeis University}
  \city{Waltham}
  \state{Massachusetts}
  \country{USA}
}

\author{Lane Harrison}
\email{ltharrison@wpi.edu }
\affiliation{%
  \institution{Worcester Polytechnic Institute}
  \city{Worcester}
  \state{Massachusetts}
  \country{USA}
}

\author{Nam Wook Kim}
\email{nam.wook.kim@bc.edu}
\affiliation{%
  \institution{Boston College}
  \city{Chestnut Hill}
  \state{Massachusetts}
  \country{USA}
}

\author{Dylan Cashman}
\email{dylancashman@brandeis.edu}
\affiliation{%
  \institution{Brandeis University}
  \city{Waltham}
  \state{Massachusetts}
  \country{USA}
}


\begin{abstract}
New tactile interfaces such as swell form printing or refreshable tactile displays promise to allow visually impaired people to analyze data.
However, it is possible that design guidelines and familiar encodings derived from experiments on the visual perception system may not be optimal for the tactile perception system.
We replicate the Cleveland and McGill study on graphical perception using swell form printing with eleven visually impaired subjects.
We find that the visually impaired subjects read charts quicker and with similar and sometimes superior accuracy than in those replications.
Based on a group interview with a subset of participants, we describe the strategies used by our subjects to read four chart types.
While our results suggest that familiar encodings based on visual perception studies can be useful in tactile graphics, our subjects also expressed a desire to use encodings designed explicitly for visually impaired people.
\end{abstract}

\ccsdesc[500]{Human-centered computing~Empirical studies in HCI}
\ccsdesc[500]{Human-centered computing~Accessibility~Empirical studies in accessibility}

\keywords{Tactile Visualization, Accessibility, Assistive Technology, Data Visualization, Graphical Perception}


\maketitle

\section{Introduction}
Data-driven decision-making is becoming increasingly ubiquitous in human experience.  Reading and interpreting data is a critical job skill, and is a fundamental method used to communicate phenomena to the broader public, most recently evidenced by the popular \textit{Flattening the Curve} campaign to slow the spread of COVID-19 in 2020~\cite{thunstrom2020benefits}.  However, the ubiquity of visual experiences may not reach blind or visually impaired (BOVI) individuals.  For example, the concept of \textit{Flattening the Curve} was primarily a visual concept, and wasn't accessible to BOVI individuals until it was translated into a tactile representation and explanation in 2021~\cite{lighthouseFlattenCurve2021}.  
Lack of access to visual representations of data can also result in a lack of access to employment.  Programs like the NSF's Data Science Corps\footnote{\url{https://new.nsf.gov/funding/opportunities/data-science-corps-dsc}}
and DARPA's data science job training initiatives  \footnote{\url{https://tools-competition.org/workforce/}} 
illustrate the importance of interacting with data in the modern job market. 
The unemployment rate for individuals with visual impairments is approximately 2-3 times the national average \cite{mcdonnall2019employment}, underscoring a significant gap in accessibility and employment opportunities. Concurrently, emerging consumer devices are aiming to bridge this gap by providing tactile representations of data, originally designed for visual perception. While tactile and visual perceptions are interconnected, particularly for individuals with residual visual memory \cite{tabrik2022}, it is unclear which findings from visual data visualization literature translate effectively into tactile formats \cite{Ault2002, Haptic2001}.

This study revisits the seminal 1984 experiment by Cleveland and McGill \cite{cleveland1984graphical}, adapting it to examine how established visual design principles translate when applied to tactile graphics for visually impaired users. While the original work assessed error in perceptual judgments of visual graphs, our replication uses tactile graphics, which is a category of graphics utilizing technologies like braille embossers and dynamically refreshing braille output devices. These technologies convert digital images into tactile formats, enhancing accessibility but prompting questions about the efficacy of traditional visual design principles in a tactile context \cite{Fritz1999, Automating2005}. Our research evaluates these principles' effectiveness across both visual and tactile media.  By understanding perceptual differences between these varying types of media, we additionally hope to unearth universal design opportunities of tactile interfaces with data for broader sets of users \cite{Non-visual2006, Teaching2006}.

In our research, we explore the following research questions:

\begin{itemize}
\item \textbf{Q1}: How does the accuracy of blind users interpreting bar charts, pie charts, bubble charts, and stacked bar charts in tactile formats compare to that of sighted users using visual formats?
\item \textbf{Q2}: How do inference times for blind users interpreting tactile graphics (bar charts, pie charts, bubble charts, and stacked bar charts) compare to those of sighted users with visual graphics?
 \end{itemize}
Based on preliminary literature reviews and our understanding of tactile perception, we have the following hypotheses.
\begin{itemize}
    \item \textbf{H1:} Visually impaired readers of tactile graphics will have the same ranking of four stimuli found in sighted participants, but the accuracy will be lower due to using different perceptual systems.
    \item \textbf{H2:} Visually impaired readers of tactile graphics will take a longer time of inference than sighted readers of visualizations because there is more processing involved, both physically through the manipulation of two hands, and mentally via comparative calculations.
\end{itemize}
Our research is a controlled experiment that directly compares the effectiveness of various chart types—bar charts, pie charts, bubble charts, and stacked bar charts—in both tactile and visual formats. Notably, it is one of the first studies to systematically compare the performance of blind and visually impaired users with that of sighted participants within the same experimental framework. By doing so, it provides empirical insights into the perceptual differences and challenges faced by visually impaired users when interacting with tactile graphics.

Through a study of eleven BOVI subjects, we find that tactile graphics can be as accurate and be read as quickly by BOVI subjects as previous studies have reported for visual graphics by sighted subjects.  We present a statistical analysis of our data showing no significant difference between sighted and BOVI subjects when compared with results from previous studies.  Through qualitative data gathered through interviews, we identify different strategies used to read tactile graphics.  We also discuss difficulties that imply both gaps in equal access to insights from data for BOVI individuals, and research opportunities for the human computer interaction research community to fill these gaps.  Our work shows that tactile graphics based on visual designs can be read accurately, but that there is also opportunity to design better encodings through human-centered design studies with BOVI participants in the future.   

\section{Related Work}

\subsection{Visual Perception Experiments}
Visual perception research in graphical interfaces began fundamentally with Cleveland and McGill’s study \cite{cleveland1984graphical}. Their foundational work laid the groundwork for understanding how people interpret graphical information. They introduced a systematic way to assess the effectiveness of different graphical representations, which has been extensively replicated and expanded upon in subsequent research. Researchers have adapted the original experiment to explore how specific audiences, such as individuals with disabilities, the elderly, and those with varying levels of visual literacy, interact with graphical data. Notable replications include Heer and Bostock’s work, which applied these principles to crowdsourced environments, showcasing how perceptions can vary widely in less controlled settings \cite{heer2010crowdsourcing}. Additionally, Talbot et al.’s focused exploration of bar chart perception provides further granularity on graphical interpretation \cite{Talbot2014}.

Recent research in visual perception experiments has increasingly focused on creating more accessible data visualizations through the integration of tactile, auditory, and multimodal feedback, addressing the needs of diverse audiences, including the visually impaired and elderly. Studies have explored the efficacy of audio narratives and haptic feedback in enhancing the accessibility of data visualizations \cite{Alexa2022, Danyang2022, WangR2022}, emphasizing the necessity for designs that accommodate non-visual modalities. Significant work has been done on developing haptic interfaces that enable blind users to interact with statistical graphics on web platforms \cite{Dajungkim2011, Darren2019}. Additionally, methodological advancements have been proposed to ensure that socio-technical considerations are integrated into visualization design \cite{Alan2019, Marriott2021}, fostering a more inclusive approach. These efforts underscore a pivotal shift towards visualizations that are not only technically proficient but also universally accessible, highlighting the critical role of inclusive design principles in the advancement of visual perception research \cite{Nihanth2022, Ather2021, ChartVi2022}.

\subsection{Accessible Computing Interfaces}

Research in accessible computing interfaces has significantly evolved, particularly focusing on data sonification and tactile feedback to enhance data interaction for users with disabilities. Studies have explored auditory feedback mechanisms that help visually impaired users understand complex data sets \cite{steven2006, Haixia2008, WangR2022, chartreader}, effectively substituting visual cues with sound. A significant focus has been on the development of sonification systems, such as the importance-driven sonification techniques of the Line Harp \cite{lineHarp2023}, which enhances line chart accessibility. Similarly, the Erie declarative grammar \cite{erie2024} facilitates data sonification, allowing users to perceive complex data through auditory means. Screen reader technologies have also seen innovative applications, notably through plugins like VoxLens \cite{Ather2022}, which enables interactive data visualization accessibility on the web. Advancements in screen reader technologies have facilitated accessible navigation through data visualizations, converting graphical elements into comprehensible auditory formats \cite{Seniz2010}. Furthermore, the integration of multimodal feedback systems combines tactile, auditory, and sometimes olfactory cues to provide a richer interaction experience \cite{dajung20211}. These efforts are not merely supplementary but are crucial for users who rely on alternative sensory channels to access information, underscoring the need for technologies that recontextualize rather than replicate data presentation across various sensory modalities \cite{Jinho2019}. This research emphasizes the importance of inclusivity in technological development and aligns with broader goals of universal design, highlighting how accessible interfaces can expand digital engagement for all users \cite{Marriott2021}. Furthermore, the Data Navigator \cite{dataNavigator2023} serves as a toolkit focusing on accessibility-centered data navigation, proving essential for users requiring non-visual data interaction. These tools not only ensure compliance with accessibility standards but also promote an inclusive approach to data engagement, highlighting the shift from traditional visual data representation to multimodal data interaction that accommodates a wider range of sensory preferences and capabilities.

\subsection{Tactile Graphics}

Tactile graphics enable BOVI individuals to interact with graphical data through embossed or raised diagrams, offering a non-visual approach to data representation. Experiments conducted by Goncu et al.~\cite{Goncu2010} highlighted preferences for tactile diagrams with gridlines and Braille values over direct transcriptions, and tactile tables over tactile charts. Engel and Weber~\cite{Engel2017} further emphasized the importance of tick marks, grid lines, texture, and the positioning of legends for the readability of tactile charts, documenting challenges like information overload and orientation issues in their survey with 71 participants~\cite{EngelWeber2017}. Additionally, they explored differences in affordances among various chart types including bar, line, scatter, and pie charts~\cite{EngelWeber2018}. Watanabe and Inaba \cite{Watanabe2018} investigated the suitable texture granularity for tactile bar charts on capsule paper through experiments where participants were asked to count the number of bars under different texture conditions. Watanabe and Mizukami \cite{WatanabeMizukami2018} conducted an experiment comparing tactile scatter plots to tactile and electronic tables, where participants were asked to identify the relationship between two variables. The results showed that tactile graphs outperformed the other two conditions. In a related study with 8 blind and low-vision participants, Yang et al.~\cite{Yalong2020} compared tactile node-link and matrix diagrams for tasks such as pathfinding and cluster identification, noting a preference for node-link diagrams, which performed best except in adjacency tasks.



\section{Methodology}

\noindent \textbf{Creation of Tactile Graphics:} 
The primary techniques for producing tactile graphics include embossing with a braille embosser, printing on swell paper, and thermoforming. Swell paper, utilized in our experiment, features microcapsules that expand when heated, creating raised features up to 0.5mm high \cite{Jonathan2003}. This method was chosen due to its precision, quick production capabilities, and cost-effective creation. The cost per tactile graphic was approximately \$2.75, making this approach economically viable for large-scale studies. An international survey of thirty blind and partially sighted people demonstrated a strong preference for swell paper, citing its enhanced tactile feedback and durability, making it ideal for users with vision impairments \cite{rowell2005feeling}. 


\begin{figure}[h]
  \centering
  \includegraphics[width=0.5\textwidth]{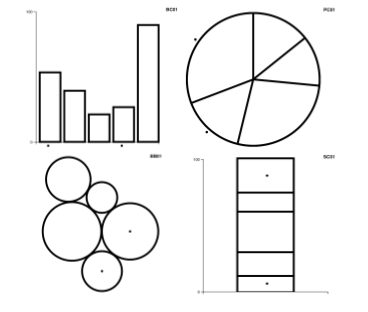}
  \caption{Examples of tactile graphics produced using the swell form machine, illustrating a 50\% data ratio across four chart types: bar, pie, bubble, and stacked.}
  \label{fig:tactile_stimuli}
\end{figure}



\noindent \textbf{Generating Stimuli:} We adapted the stimuli generation process from the original Cleveland and McGill study to accommodate tactile perception. This involved constraining the ratios of the smaller to the larger graphical elements, which we refer to as the \textbf{True Proportional Difference}, uniformly in 5\% increments from 50\% to 95\%, while allowing the larger elements to vary from 40\% to 100\% of the viewport. This variance was intended to provide a broad spectrum of graphical sizes, ensuring the tactile graphics were legible and effectively represented the data. Distractor stimuli were also included, ranging from 20\% to 100\% of the viewport, to mimic real-world scenarios where multiple data points are presented together.  In total, 80 different charts were produced, with two variations for each of the ten different values of True Proportional Difference (50\%, 55\%, \ldots, 90\%, 95\%), and four chart types.  We recognize that previous versions of this experiment used different values of True Proportional Difference, notably lower than 50\%.  Out of concerns for cost, we were limited to visualizing data that could easily be represented in swell form paper. For this purpose, we compare not only overall error but midmeans stratified by True Proportional Difference.  


\section{Experimental Design}

\noindent \textbf{Participant Recruitment and Setup:} 
In the first round of interviews, eleven participants in total were recruited from a local school for the blind and the experiments were conducted over two days in a controlled environment. The participants ranged in age from 18 to over 65, with an average age of approximately 48.3 years. The gender distribution included 6 males and 5 females. Education levels varied, with 5 participants having completed an undergraduate degree and 5 holding a master's degree, while 1 had some graduate education. Each session lasted approximately 75 minutes and was conducted in person to enable effective interaction with the tactile graphics. Two months later, a group interview was held with four of the participants of the original study to discuss strategies and to photograph interactions with the tactile graphics.

\noindent \textbf{Procedure:} The procedure began with an informed consent briefing read aloud to participants, followed by a practice session with eight stimuli (two of each chart type) to familiarize them with the tactile graphics. Participants were then presented with the remaining 72 graphics in a randomized order and asked to interpret these at their natural pace, verbalizing their thoughts and answers for audio recording.  Participants were told the true answer (percent size of the smaller stimuli to the larger) for their 8 training stimuli, but were not told during the experimental phase.  Answers were recorded into a spreadsheet during test time and then later adjudicated against recordings, with any disagreements removed from the dataset.  This phase of the experiment was limited to 30 minutes, and participants were notified when half of the time was left.

\noindent \textbf{Data Collection:}  Data were collected through both quantitative and qualitative methods. Quantitatively, we measured the accuracy of each response and the time taken to complete each task. Qualitatively, participants were asked to provide feedback about their experience, including any difficulties they encountered and their subjective assessment of the tactile graphs’ clarity and usability.  After the tactile interpretation tasks, participants completed a demographics survey and engaged in a 10-15 minute semi-structured interview. These interviews aimed to gather qualitative feedback on their experience and any challenges they faced during the experiment.  \footnote{Study documents, including interview questions, demographic questions, and graphics files are available at \url{https://osf.io/3nsfp/?view_only=7b7b8dcbae1d4c9a8bb4325053d13d9f}.}

\noindent \textbf{Followup Group Interview:} Two months after the experiment was conducted, participants were contacted via email to participate in a followup group interview in order to compare strategies for reading the tactile graphics.  Four participants joined the experiments in a conference room for a 90 minute group interview.  For each of the four chart types, the participants were first each given two examples of the chart type (one close to a ground truth of 50\%, one close to a ground truth of 95\%) in order to refresh their memory about the chart type.  Then, they were each asked about their strategies for reading from the chart, what they found difficult about reading the chart, and any improvements they suggested.  The group interview ended by asking the four participants to respond to any of the strategies of the other participants for any of the chart types.  The meeting was photographed and its audio was recorded and transcribed for analysis of common strategies.

\begin{figure*}[h]
        \subfloat[Bar Charts]{%
            \includegraphics[width=.25\linewidth, height=0.8in]{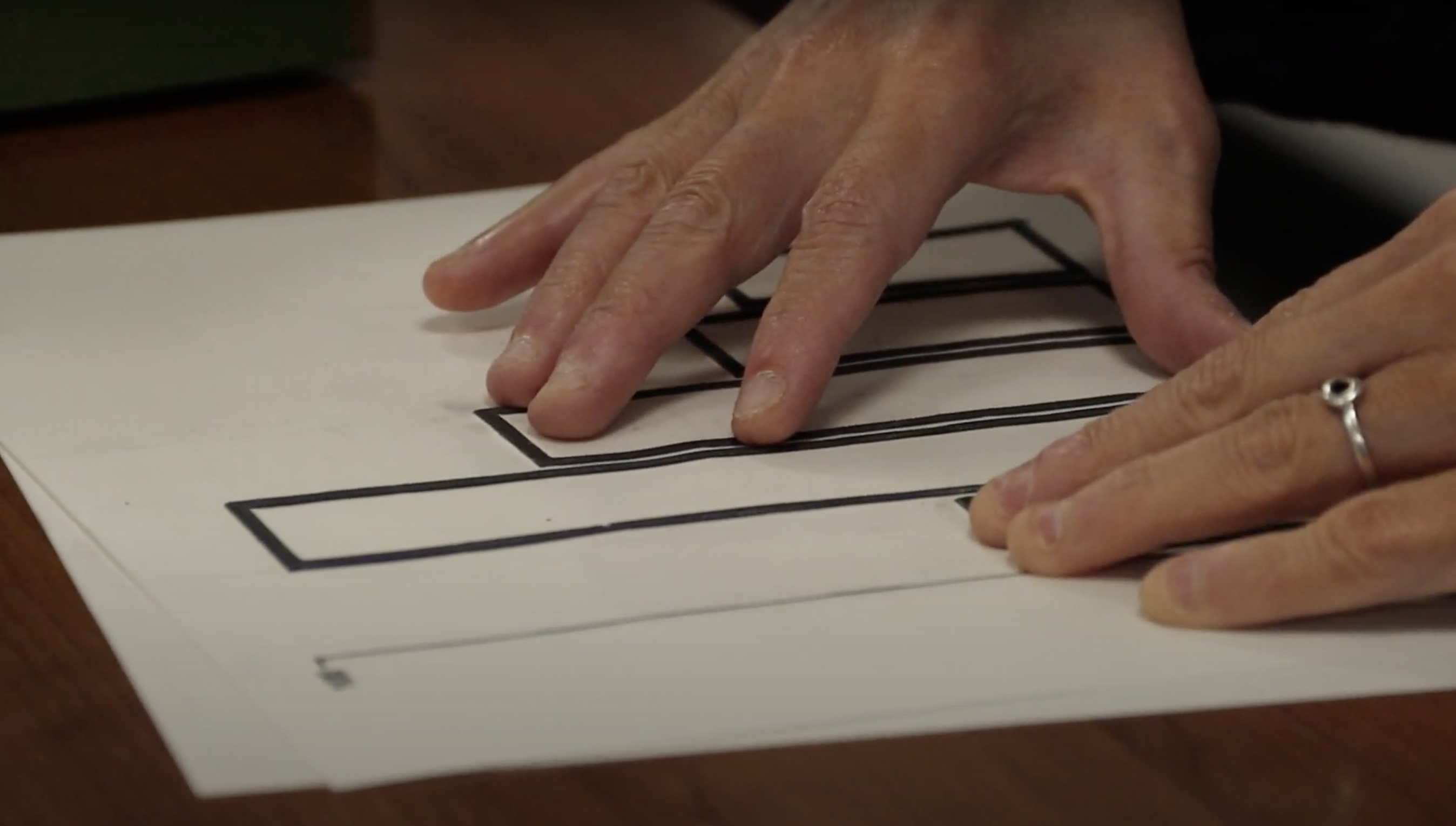}%
            \label{subfig:a}%
        }\hspace{2pt}
        \subfloat[Pie Charts]{%
            \includegraphics[width=.25\linewidth, height=0.8in]{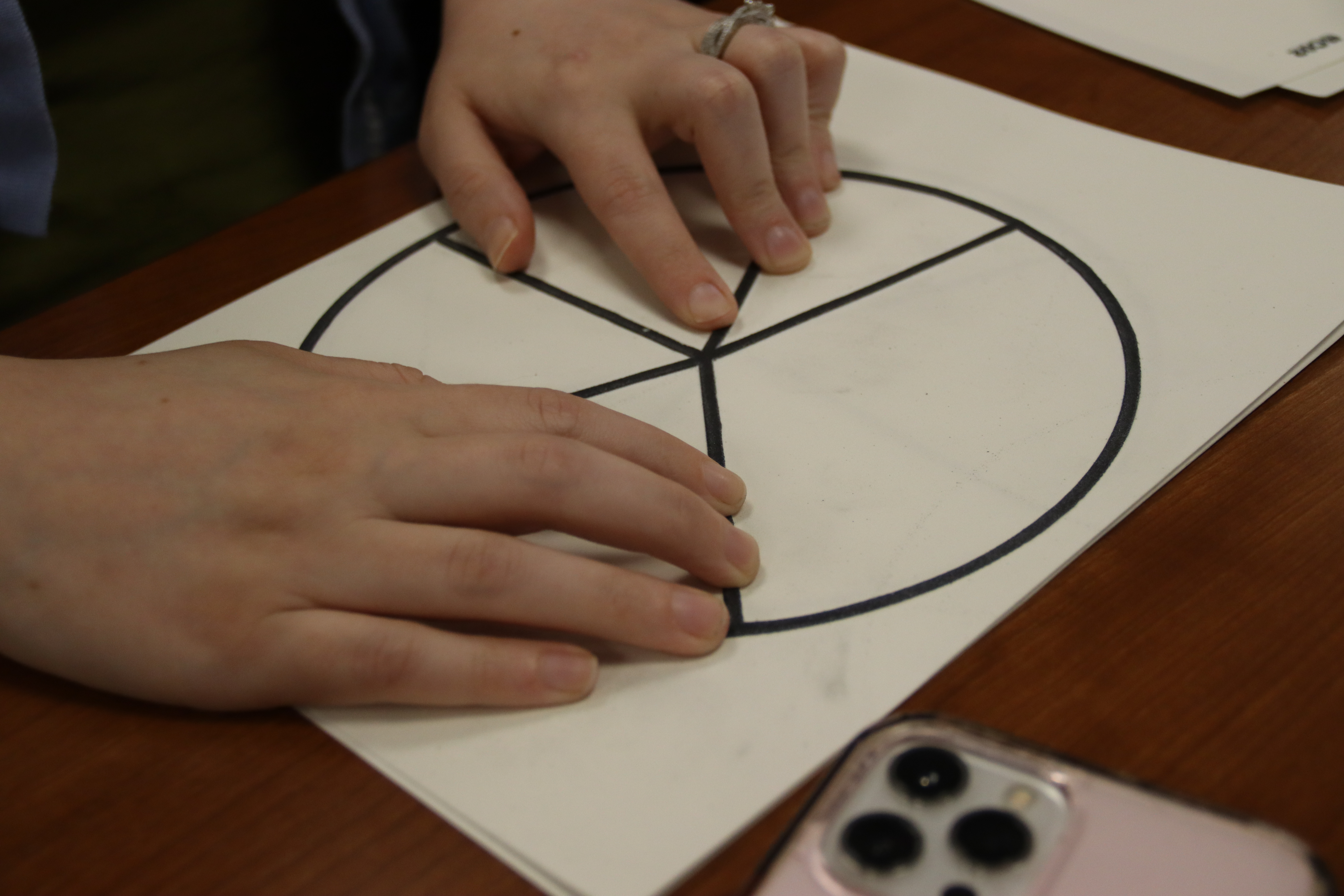}%
            \label{subfig:b}%
        }\\
        \subfloat[Stacked Bar Charts]{%
            \includegraphics[width=.25\linewidth, height=0.8in]{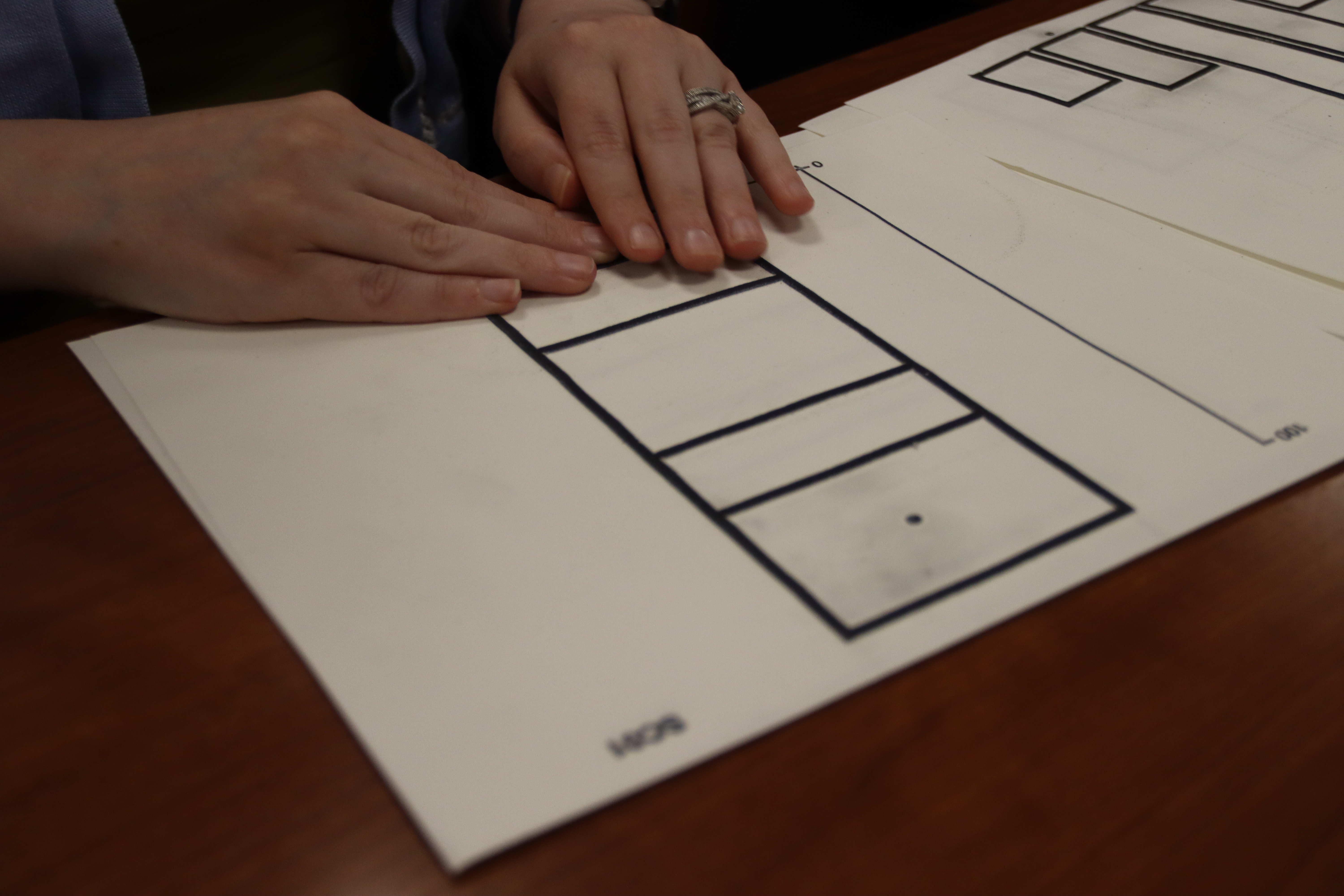}%
            \label{subfig:c}%
        }\hspace{2pt}
        \subfloat[Bubble Charts]{%
            \includegraphics[width=.25\linewidth, height=0.8in]{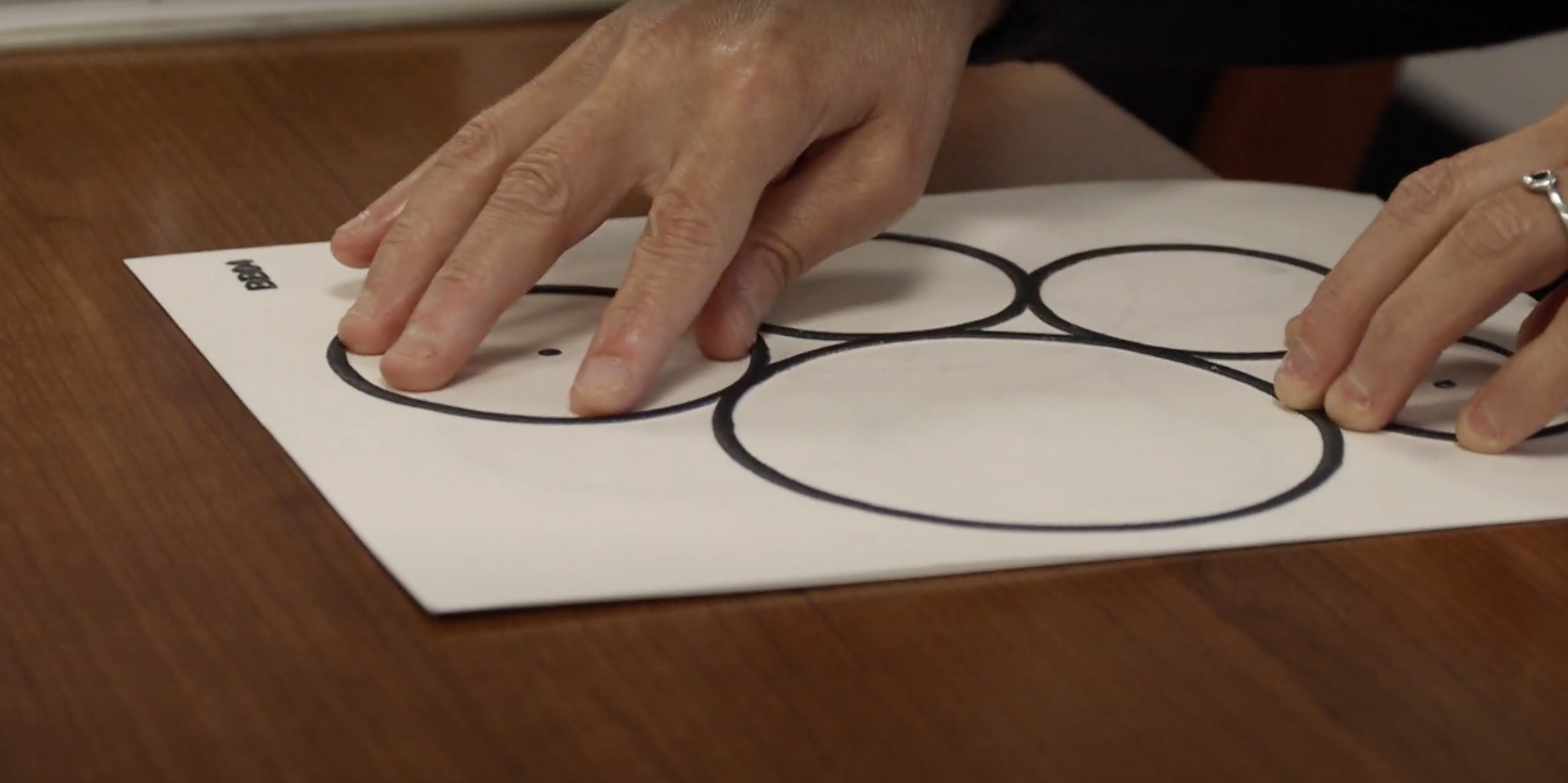}%
            \label{subfig:d}%
        }
        \caption{Participants interacting with different types of tactile graphics during the follow-up group interview: (a) Bar Chart, (b) Pie Chart, (c) Stacked Bar Chart, (d) Bubble Chart.}
        \label{fig:fig}
    \end{figure*}

\section{Results}
\label{sec:supplement_inst}

\begin{figure}[h]
 \centering 
 \includegraphics[width=\columnwidth]{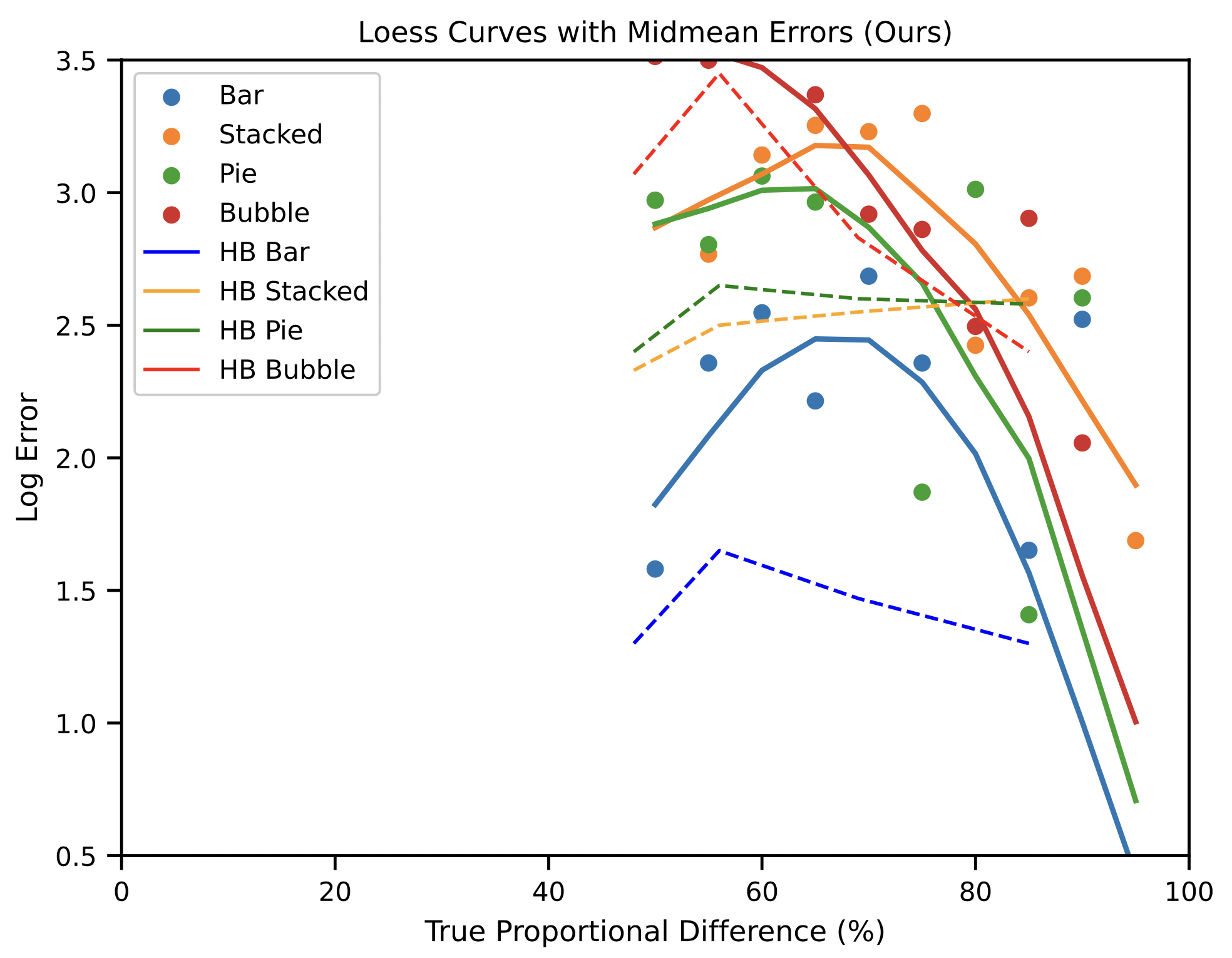}
 \caption{Midmean errors by chart type for ours (solid lines) and Heer and Bostock (2010, sighted participants).  Both datasets show the same general trend, but midmean errors are lower in Heer and Bostock for bar charts and stacked bar charts, a reversal of the overall errors in Figure~\ref{fig:error_rates}, suggesting outliers may have had an effect in our study.  }
 \label{fig:midmeans}
\end{figure}

\begin{figure}[h]
 \centering 
 \includegraphics[width=\columnwidth]{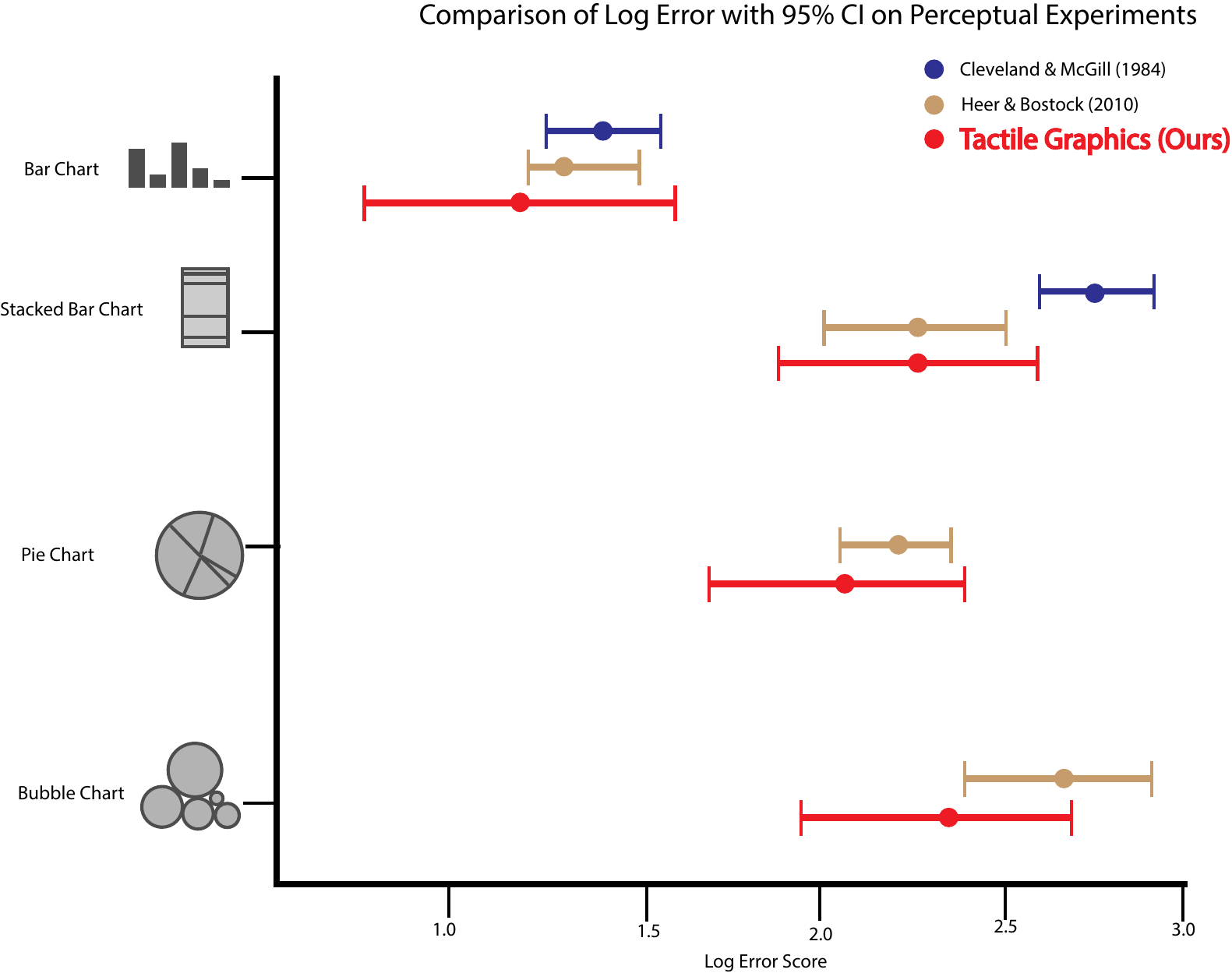}
 \caption{Error scores for our study (red) compared with two previous comparable studies: Cleveland and McGill~\cite{cleveland1984graphical} and Heer and Bostock~\cite{heer2010crowdsourcing}. Error bars indicate 95\% confidence intervals via bootstrapping.  Pie Chart and Bubble Chart were not measured in Cleveland and McGill.  Previous studies' results were derived via published graphics rather than raw data.}
 \label{fig:error_rates}
\end{figure}



In accordance with the original Cleveland And McGill study, we calculated both midmeans for absolute accuracy by ground truth per chart and means and 95\% confidence intervals of error scores for each four bar charts.  Confidence intervals are calculated through bootstrapping.  Error is scored as the log of the absolute difference plus $\frac{1}{8}$, which is added for numerical stability.

\[
\text{Error Score} = \log\left(\left|\text{guessed ratio} - \text{true ratio}\right| + \frac{1}{8}\right)
\]

The error scores, seen in Fig~\ref{fig:error_rates}, match up closely with previous results on sighted individuals both in laboratory settings and through a crowdsourced replication on Amazon Mechanical Turk~\cite{cleveland1984graphical, heer2010crowdsourcing}.  Our results have larger confidence intervals, which is partly explained by a smaller number of trials in our study ($n=651$) vs. the other two  ($n=3481$ for the crowdsourced and $n=2550$ for the laboratory experiment).  However, there are other possible explanations; we analyze the between-subject variance as a source of uncertainty in the subsequent section.

The midmeans chart (Fig~\ref{fig:midmeans}) displays the middle quartiles of error for each group of ground truth, and was used in previous studies to communicate performance without being influenced by outliers.  It indicates that the bar chart is the most accurate chart in general.  Across all chart types, the error decreases as the true proportional difference gets close to 100\%, with a likely peak in error near 50\%.   This matches previous findings reported in a crowdsourced study with sighted participants~\cite{heer2010crowdsourcing}.  However, while the ranking of different charts is similar, there are some differences: the performance on the stacked bar and bar charts had higher midmean error in our study.  This conflicts with the mean error in aggregate seen in Fig~\ref{fig:error_rates} - it appears that the removal of outliers from the midmean calculations resulted in our study's error being higher than the crowdsourced study.

\subsection{Analysis of Hypotheses}

Our first hypothesis, \textbf{H1}, was that BOVI readers of tactile graphics would have the same ranking of the four chart types found in sighted participants, but that the accuracy would be lower.  Our data provided only partial evidence for \textbf{H1}. Both midmeans and error confidence demonstrated similar rankings of charts were similar, but in aggregate the accuracies of BOVI readers was similar or even higher in some cases, as with the bar chart.  One-sided t-tests could not confirm the hypothesis that the error of BOVI readers was higher than previous studies ($p>0.5$ in comparisons with Cleveland and McGill~\cite{cleveland1984graphical} and Heer and Bostock~\cite{heer2010crowdsourcing}).  


Our second hypothesis, \textbf{H2}, was that BOVI readers of tactile graphics would take a longer time of inference than sighted readers of visualizations due to the physical and mental processing required.  We compare our results from those reported in Heer and Bostock.  Our participants, on average, viewed 59 charts making the total number of readings $n=651$. The average completion time per chart judgment was 26.74 seconds for BOVI users interacting with tactile graphics. In contrast, Heer and Bostock's MTurk study reported a higher average time of 54 seconds per trial, with a median response time of 42 seconds and a standard deviation of 41 seconds.  A one-sided t test finds this data does not support \textbf{H2}.  This comparison suggests that our tactile graphics, while designed for BOVI users, enabled relatively faster responses compared to the MTurk study.  We note that Heer \& Bostock suggest that in a laboratory setting, compared with MTurk, they expect that participants would be faster. 

\subsection{Hierarchical Analysis}

Our results indicated some nuance in the data that was not captured by the first two hypotheses.  In particular, the midmeans chart and the aggregate error disagreed on whether the performance of our subjects was better, equivalent, or worse, than results from previous replications of this experiment with sighted participants.  A recent work by Davis et al. offers a different type of analysis of graphical perception studies that enables statistical statements about the variance between subjects using hierarchical modeling~\cite{davis2022risks}.  We repeat that analysis with our results to better understand the outcome of our experiment.

To fit our data to a hierarchical model, we need to declare the assumptions of the model.  Our assumptions will model how the data was generated, parameterizing the various effects that we believe led to the values offered by our subjects.  We assume that these effects are stochastic, and so we model them as random variables, resulting in a Bayesian model.  We then use numerical methods to determine the likely posterior distributions for these variables when conditioned on the data generated by our experiment.  For a broader overview of this technique, see Gelman et al.~\cite{gelman2020bayesian}

Davis et al. build a hierarchical model by starting with a simple linear model, which assumes that the average performance of each participant is normally distributed about the true mean for the population - essentially that any differences between participants are the result of random chance.  They then expand their model by changing different assumptions.  These assumptions include restrictions on the output of the model (replacing the normal distribution with the Zero-Inflated beta distribution), an additional \textit{random effects} term that assumes some individual differences effect between participant and visualization ($U_\mu[\text{vis}[i], \text{participant}[i]]$), and additionally learning submodels for the precision ($\phi_i$) and probability of zeros($\pi_i$).  

\begin{flalign*}
\text{abs\_error}[i] &\sim \text{ZeroInflatedBeta}(\mu[i], \sigma[i], \pi[i]) & \textit{likelihood}\\
\text{logit}(\mu[i]) &= \beta_\mu[\text{vis}[i] + U_\mu[\text{vis}[i], \text{participant}[i]] & \textit{mean submodel}\\
\text{log}(\phi[i]) &= \beta_\phi[\text{vis}[i] + U_\phi[\text{vis}[i], \text{participant}[i]] & \textit{precision submodel}\\
\text{logit}(\pi[i]) &= \beta_\pi[\text{vis}[i] + U_\pi[\text{vis}[i], \text{participant}[i]] & \textit{zeros submodel}\\
\end{flalign*}

By adding these additional terms and adding submodels, in sum, the hierarchical model is able to capture a participant having different types of error or bias for each visualization type.  We replicate the same model used in that work, including using the same weakly-informed priors as they reported, and then compare our analysis to their study.  We note that the number of participants (eleven vs. more than a hundred) and the number of trials per participant (approximately sixty vs. more than a thousand) results in experimental datasets of different scales, and so some of the analysis of the resulting hierarchical models still had high levels of uncertainty.  But we highlight two charts that can tell us more about the difference between the BOVI subjects of our study and the sighted participants in the previous study.

\begin{figure}[h]
 \centering 
 \includegraphics[width=\columnwidth]{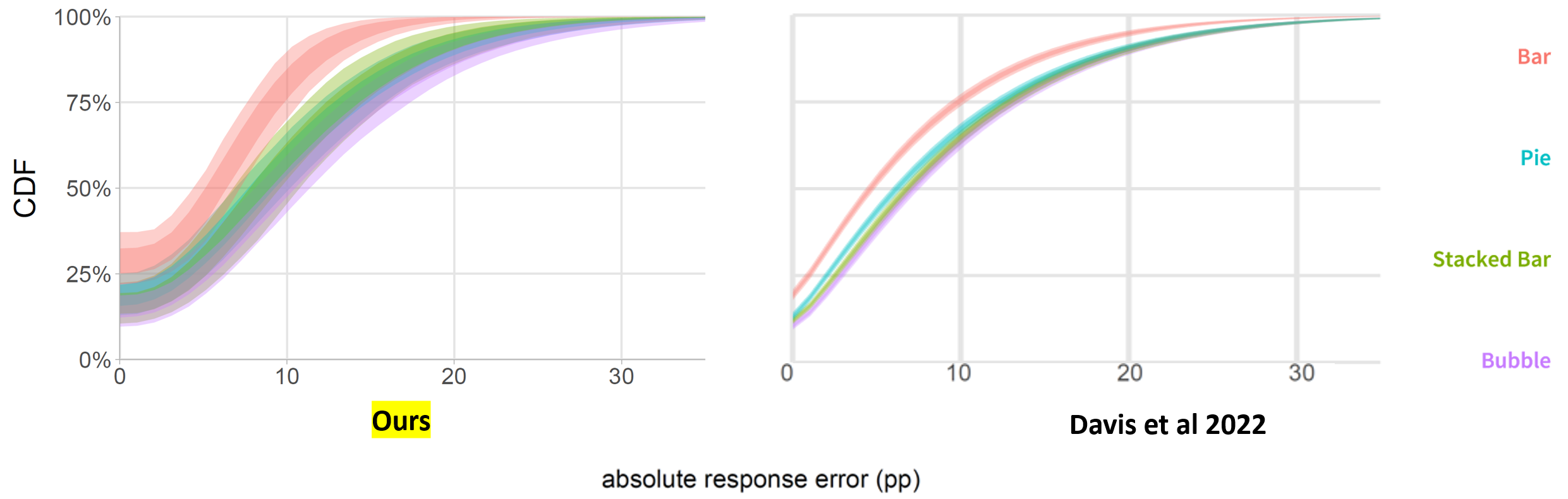}
 \caption{Cumulative Distribution Functions (CDFs) of the hierarchical model comparing our data (left) to data from Davis et al (right)~\cite{davis2022risks}.  Davis et al. introduced this model to allow for additional effects between participants and visualization types and to learn distributions of error.  This plot illustrates that errors within our experiment had similar rankings to Davis et al (\textbf{H1}), but the difference in shapes suggests that our study resulted in fewer small error estimations, indicated by the flat left side of the curves in our plot. }
 \label{fig:cdf_ours}
\end{figure}

First, in Fig~\ref{fig:cdf_ours}, we compare the Cumulative Distribution Functions of the hierarchical models of the absolute response error of the participants' answers.  We are able to produce this chart with the hierarchical model from Davis et al. because it calculates a posterior distribution rather than point estimates, letting us look at the shape of the types of error we found in our study.  It is notable that the both studies have both the same ranking of chart types consistently through different levels of error, both low (less than 5 absolute error) and high (more than 20 absolute error).  At the same time, our study appears to have less errors with very small error (less than 5 absolute error), indicated by the flat region at the very left of the plot.  This is consistent with repeated think aloud statements by our subjects that it was very hard to decide on small differences of estimated answers, i.e. 70\% vs. 75\%.  

\begin{figure}[h]
 \centering 
 \includegraphics[width=\columnwidth]{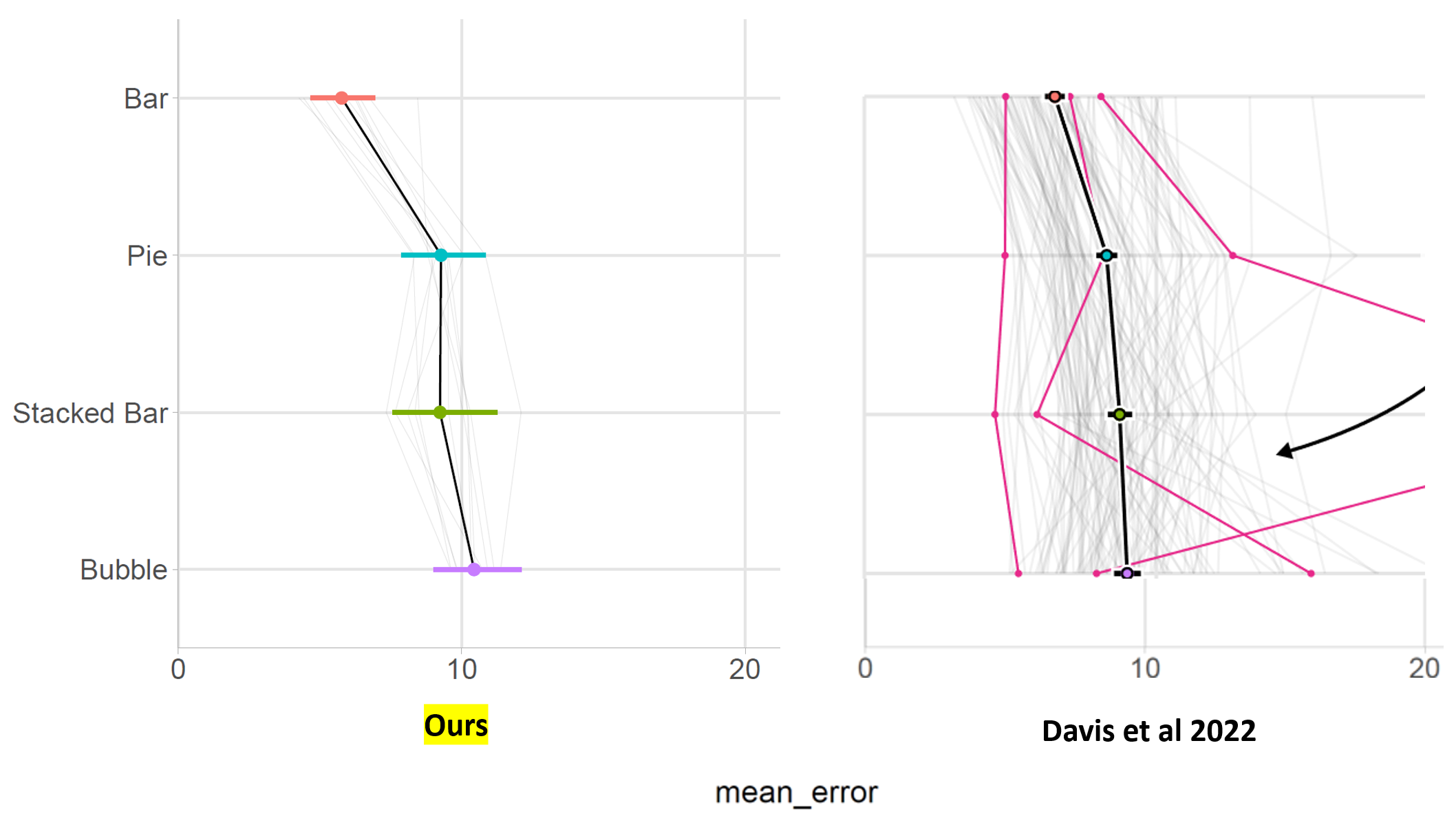}
 \caption{A parallel coordinates chart showing the mean absolute error of our eleven participants, based on a chart from Davis et al~\cite{davis2022risks}.   Results indicate that the performance varies greatly between the most accurate and least accurate participants within our study.  However, it does not seem to be significantly different than the variance in performances illustrated in Davis et al, suggesting additional studies with larger participant pools are needed to explore the between-subject variance.}
 \label{fig:parallel_coords}
\end{figure}

Next, in Fig~\ref{fig:parallel_coords}, we compare the mean absolute error across our eleven participants on each chart type with errors from the Davis et al. experiment.  In the parallel coordinates chart, the mean performance of each participant across all 10-20 trials from each of the four chart types is calculated.  This chart was used in Davis et al. to show that some participants have different rankings or relative performances across visualization types - one pink line, for example, showed a bar chart being less effective than a stacked bar chart.  In our data, on the left, we do see similar variance in the rankings of the four chart types within our eleven participants.  It is also notable that the performance appeared to vary broadly in our sample of eleven participants.  However, it isn't clear that the variance noticed in our study was markedly different than the variance evident in Davis et al.

Our hierarchical analysis revealed both similarities and differences with previously reported results on a wider study run on sighted participants within a crowdsourced study.  By fitting the data to a Bayesian model we are able to analyze the CDFs and note the the performance of BOVI participants is similarly distributed across true error types, although there is some difference in the distribution of guesses with small error.  In addition, the comparison of mean performances per participant per chart type showed that while our participants had noticeable variance in performance and preference of charts, it wasn't clear if it was different than previously reported results.  Both findings suggest that a larger study would be needed to analyze the types of between-subject and within-subject variance that appears in the data, and to compare it with statistical significance to previous studies.

\subsection{Qualitative Analysis of Subject Strategies}

Through think-aloud statements during the experiment, ad-hoc interviews after the experiment with all eleven participants, and the followup group interview with four participants, we gathered qualitative data about the difficulties encountered by our subjects in reading our tactile graphics, as well as the strategies they employed.  We conduct a thematic analysis of this data, clustering statements into themes and highlighting quotes from our subjects.

\subsubsection{Strategies}  

While the strategies varied slightly between chart types and between participants, there were two broad categories used to measure distance-based encodings, such as the length of a bar or the length of an arc of a pie chart wedge (\textbf{calipers} and \textbf{rulers}), and one strategy for measuring areas (\textbf{splaying fingers}).  

\begin{figure}[ht!]
    \centering
    \begin{minipage}{0.45\textwidth}
        \centering
        \includegraphics[width=\linewidth, height=0.6\linewidth]{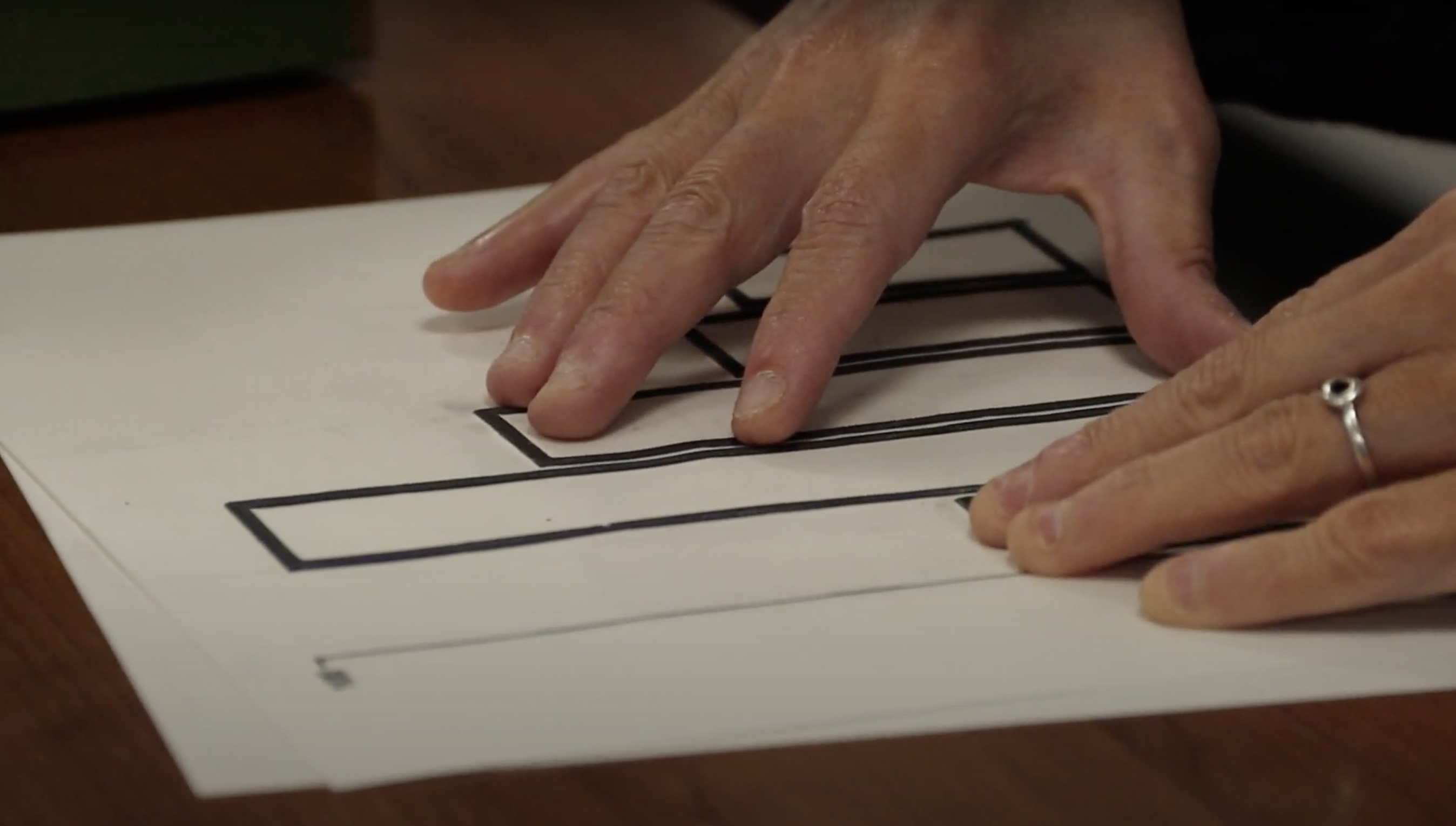} 
        \caption{Caliper Strategy: A participant using thumb and forefinger as calipers to measure distances, locking the measurement and moving to compare with another.}
        \label{fig:caliper_strategy}
    \end{minipage}\hspace{10pt}
    \begin{minipage}{0.45\textwidth}
        \centering
        \includegraphics[width=\linewidth, height=0.6\linewidth]{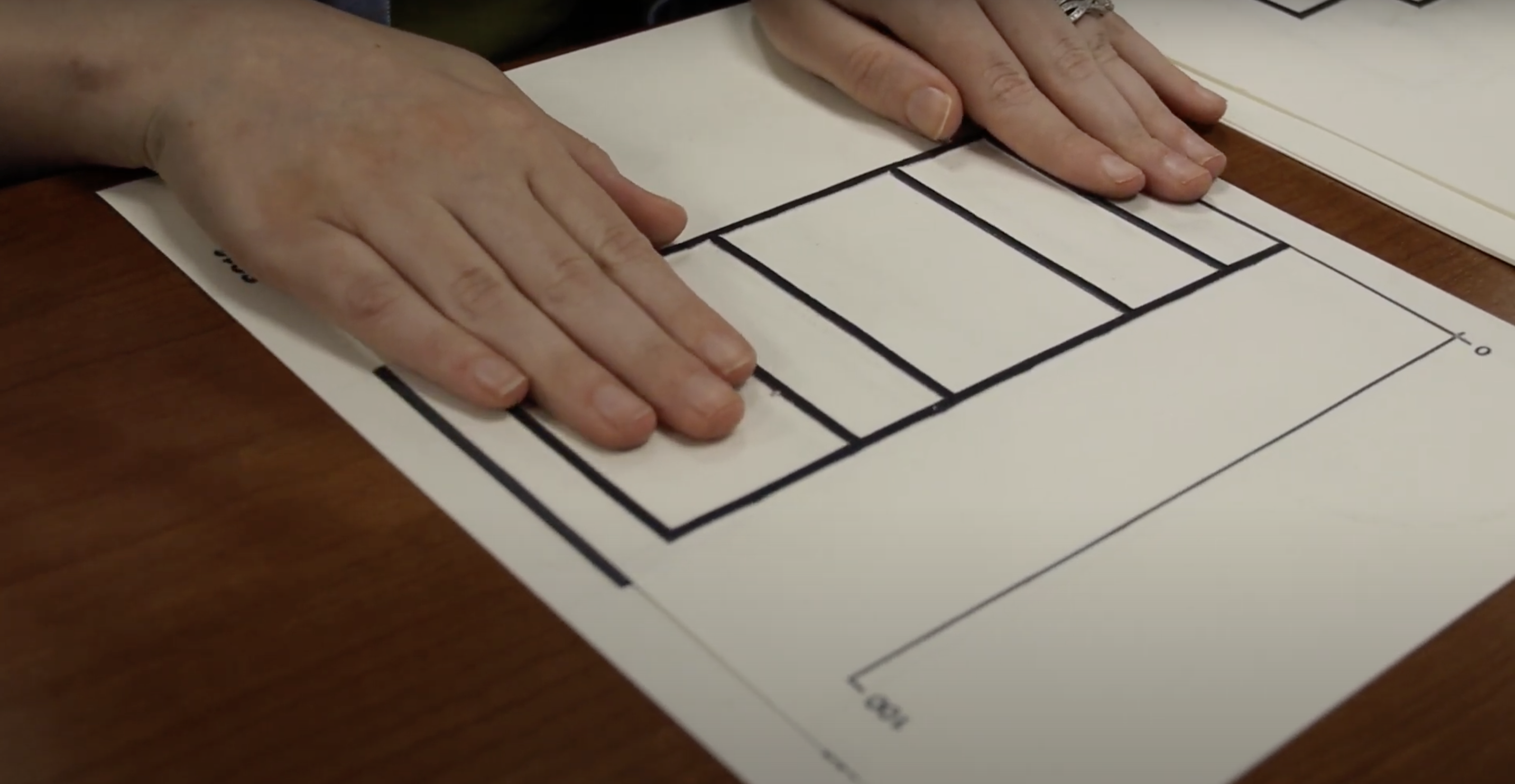} 
        \caption{Ruler Strategy: A participant using the widths of different fingers to estimate lengths of bars and compare measurements on tactile graphics.}
        \label{fig:finger_measure_strategy}
    \end{minipage}
\end{figure}

In the first strategy, the thumb and forefinger are used to measure a length between two risen bumps in the tactile graphics (see Fig~\ref{fig:caliper_strategy}).  In a comparison task like in our experiment, the hand that makes this first measurement locks, similar to a caliper, and then is slid over to the second measurement (i.e. moving from the smaller bar to the larger bar).  The second measurement was typically made using a second hand also as a caliper.  The two calipers were then described as being compared, or even visualized, to estimate what the relative measurement between them would be.  

In the second strategy, the fingers of one hand were used as measurements of length, i.e. one bar was a pinkie, ring finger, and middle finger, while the other bar was just a pinkie and a ring finger (Fig~\ref{fig:finger_measure_strategy}).  In this strategy, participants often pointed out that it was difficult when the measurement didn't match perfectly with one of their fingers (i.e. the length was two and a half fingers), and that it was also difficult translating from finger widths to relative percentages, since fingers are of different lengths.  

\begin{figure}[ht!]
    \centering
    \includegraphics[width=0.45\textwidth, height=0.3\textwidth]{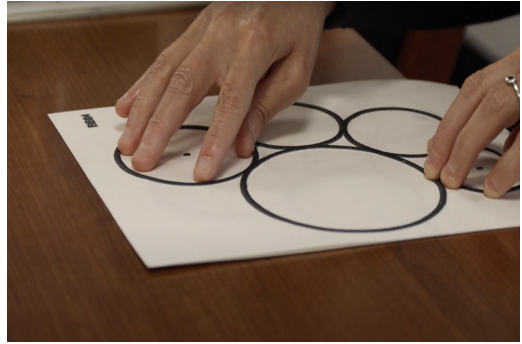} 
    \caption{Splaying Finger Strategy: A participant use the splaying of fingers to measure and interpret areas in tactile graphics.}
    \label{fig:splaying_finger_method}
\end{figure}

While all four chart types can be read via measuring the length of a feature of the graphical primitive, our participants more frequently measured the full area of bubble charts, and sometimes measured full area of the wedges of the pie chart.  The general strategy for measuring area involved locating the center of the area, and then splaying fingers outward to sense the size of the area (Fig~\ref{fig:splaying_finger_method}).  Some participants would use both hands to measure areas, while others would use the same hand to measure both stimuli to make their relative measurements.  This was less effective because it doesn't rely on a static measurement, like the calipers or ruler method, that can be shifted from one stimuli to the next.  This may explain why area marks like those in the bubble chart result in greater error in our experimental data.

For stacked bar charts and pie charts, participants frequently rotated the chart to make the physical measurements more comfortable from their seating position.  An example can be seen in the participant's reading of the stacked bar chart in Fig~\ref{fig:finger_measure_strategy}, where the fingers pointing away from the body should be perpendicular to the axis being measured.  The orientation of the graphical primitives appears to be important for tactile perception in a unique way compared to visual graphical primitives, and should be analyzed in a future study.

Lastly, participants frequently made multiple readings of the same chart to check their work.  This was described by a participant as necessary since they were making only estimations of the encoded values, and so by estimating twice, they have less of a chance of making a mistaken estimate.  It is not known whether visual perception involves multiple redundant measurements of the same visual channels because the measurement can be implicit, but it should be recognized that any primitives designed for tactile graphics should account for this need to redundantly measure to mitigate uncertainty in estimates.

\subsubsection{Difficulties}

Participants described their frustrations throughout think aloud, and were explicitly asked about what made it difficult to read the  charts.  We broadly categorize these difficulties into challenges with \textbf{primitive scale}, \textbf{distance traveled}, \textbf{tactile noise}, \textbf{local vs. global} sense of the data, and \textbf{visual memory}.   For scale, we note that both the calipers strategy and the wruler strategy have limitations in the scale of graphics they can measure.  While visualizations for sighted viewers have limitations of scale as well, reasonable scales may be much larger (i.e. large displays) and much smaller (i.e. Apple Watch), than the reasonable scales for tactile graphics.  Participants mentioned that having smaller hands felt like a disadvantage for some of the larger graphical primitives being read.  The amount and orientation of the distance between marks being measured also added additional difficulties, particularly with the caliper strategy, which often required the measuring hand to move from one mark to the other to make a comparison.  The longer this distance was, the greater chance of accidentally shifting the measurement of the caliper.  This was sometimes exacerbated by marks that had orientations that were not axis-aligned, as in the bubble chart and pie chart.

Starting from Tufte's rules about maximizing the data-ink ratio~\cite{tufte1983visual}, it is accepted that additional noise in the visual encoding can distract from the underlying perceptual task.  Our participants mentioned that the design of our graphics contained noise that made it difficult to focus on or even locate the particular marks being measured.  However, the types of noise that were identified were not anticipated.  First, participants noted that the width of the lines in our swell form graphics were disruptive, and that they would prefer thinner lines.  It was also noted that the white space between marks was possibly misleading, because the positive space inside the marks and the negative space between marks may be harder to discriminate in tactile perception.  The bar chart was particularly challenged by white space, since the space between marks is fully redundant.  Lastly, because participants can only use part of their hand at once and do not interact with the entire marks, the marks themselves may not have needed their full tactile encoding.  This was most prevalent in the bubble chart, where the areas where the circles touch are the most noisy areas, but those areas were not typically used by participants.  It may be that space-filling visualizations like bubble charts are not effective for tactile visualizations because of the noise resulting from the cramped and unaligned layout of marks.  In general, the heuristics and gestalt rules~\cite{wong2010points} taken for granted in visual design should be interrogated for tactile design.

A broader difficulty that was experienced by our participants was the relationship between a \textit{global} view of the data and a \textit{local} view of the data.  Shneiderman's mantra (\textit{overview first, zoom and filter, details on demand}, a motivating heuristic for visualization design, states that a global view of the data is usually the first impression that a user wants~\cite{shneiderman2003eyes}.  However, one participant brought up independently that they did not typically generate a global perception of the data: \textit{"When you're using your fingers to do things, your hand is covering things you can't be viewing.  We can't look at these things globally.  We have to look at them in a micro- rather than a macro- kind of way."}.  This suggests that elementary perceptual tasks like those used in this study and Cleveland and McGill, which are typically local, might be expected to be comparable between sighted and visually limited participants, but more global tasks could result in a wider gap and greater need for new encodings for tactile graphics.

Lastly, participants frequently brought up the significance of having visual memory for perceiving graphical primitives.  They described that it is generally well known within the BOVI community that those who become blind during their lives often have visual memories that can improve their ability to interpret visual concepts.  One participant of our group interview was born blind, rather than becoming blind during their life, and remarked at how they found the process of reading tactile graphics very frustrating: \textit{"I've never seen a graph, I've never seen anything visually.  I may be frank -- graphics mean nothing to me.  I have no context, so I would dismiss them out of hand if \textbf{they aren't built for me}...  If they're built from a visual place, at least for me, they mean nothing, they have no resonance with me whatsoever}.  This suggests that additional studies are necessary to understand whether the presence of visual memory impacts the perceptual accuracy of BOVI subjects.  It also suggests that a user-driven design of tactile graphical primitives designed primarily for those born blind could result in more effective encodings for tactile graphics.









\section{Discussion and Future Work}

\subsection{Future of Tactile Graphics}
Our findings suggest that while tactile graphics adapted from visual design principles can be beneficial, they may not fully meet the accessibility needs of those born without sight. This aligns with recent studies emphasizing the importance of designing with and for BOVI users to create truly effective and accessible tactile graphics \cite{zong2022rich, lundgard2021accessible}. As \cite{reinders2024refreshable} suggests, incorporating conversational agents and advanced tactile displays could further enhance accessibility, pointing towards a multimodal approach as a promising future direction. Our results support this direction, indicating that tactile graphics should be part of a broader, inclusive design strategy that prioritizes user-centric development.

\subsection{Limitations}
While it produced useful insights, our study had some limitations. The primary constraint is the reliance on existing visual encodings which might not optimally translate into tactile formats. This limitation is evident in our mixed results where some tactile graphics performed well while others did not, indicating a possible mismatch in the tactile adaptation of visual data \cite{kim2023exploring}. Additionally, our sample size and the diversity of our participant group, though adequate for preliminary insights, may not fully represent the wider BOVI population, possibly limiting the generalizability of our findings.

The design of tactile graphics itself, produced using swell paper technology, introduces another limitation due to the restricted resolution and detail this method can provide. This potentially affects the users' ability to discern fine details in complex graphs, potentially skewing the accuracy and speed of data interpretation when compared to visual graphs as noted in foundational studies \cite{cleveland1984graphical}. The accessibility and cost of producing high-quality tactile graphics also pose significant limitations. While swell form technology is cost-effective, it is not universally accessible, and the financial burden may be prohibitive for some institutions or individuals, limiting widespread adoption.

Moreover, the diversity of techniques used by participants to interpret the data underscores a limitation in the standardization of tactile graphics. For example, one participant employed a method of using fingers as calipers to measure the length of the bars, while another estimated spatial divisions through mental segmentations. Such personalized techniques, although creative, highlight the lack of a one-size-fits-all approach in the current tactile graphic designs.

Participants also noted difficulties with the texture and construction of tactile elements. Many found the uniformity in the texture of bars confusing and suggested that varying the texture could aid in differentiation. Double encoding, which combines tactile information with other sensory cues like varied textures, was recommended to improve usability and accuracy.

\subsection{Future Studies}
Looking forward, we propose a series of large-scale studies to further validate and refine the tactile graphic designs. These studies should explore a wider array of tactile and multimodal presentation techniques, involving more diverse participant groups to enhance the reliability and applicability of the findings. Furthermore, collaboration with BOVI participants should be emphasized to tailor designs more closely to user needs, potentially involving technologies that integrate tactile feedback with auditory and possibly olfactory cues for a richer user experience \cite{zong2022rich}. We anticipate that such inclusive research efforts will be crucial in addressing the broader accessibility challenges faced by the aging population with vision loss \cite{visionserve2022}.

\section{Conclusion}
In this research, we replicated the Cleveland and McGill study on graphical perception using tactile graphics to explore their efficacy for blind or visually impaired individuals. Our findings suggest that while tactile representations derived from visual design principles hold promise, their efficacy is not universally optimal, particularly for those born without sight (e.g., \cite{reinders2024refreshable}). This highlights the need for designs that explicitly consider the unique perceptual requirements of blind or visually impaired users.

The exploration of tactile graphics in our study suggests that they can serve as effective tools for data representation for the visually impaired when adapted with consideration for tactile perception. However, the mixed results across different chart types indicate that further refinement is necessary to fully leverage these tools. This aligns with recent advances in multimodal data representation which suggest integrating tactile feedback with auditory and possibly olfactory cues to enrich the data interaction experience \cite{lundgard2021accessible,zong2022rich}.

Future research should continue to innovate in the design of accessible data visualizations by engaging with blind or visually impaired users throughout the design process. This user-centric approach is essential for developing effective tactile graphics that are truly accessible and useful in practical scenarios. Further large-scale studies involving a broader spectrum of tactile and multimodal presentation techniques are recommended to enhance the reliability and applicability of these technologies.

As we look towards the future, it is clear that tactile graphics will continue to evolve, reflecting the advancements in technology and a deeper understanding of accessible design. It is our hope that these efforts will significantly reduce the barriers faced by the visually impaired community, granting them greater access to the burgeoning field of data science and visual analytics.

\bibliographystyle{ACM-Reference-Format}
\bibliography{ref}


\begin{thebibliography}{52}


\ifx \showCODEN    \undefined \def \showCODEN     #1{\unskip}     \fi
\ifx \showDOI      \undefined \def \showDOI       #1{#1}\fi
\ifx \showISBNx    \undefined \def \showISBNx     #1{\unskip}     \fi
\ifx \showISBNxiii \undefined \def \showISBNxiii  #1{\unskip}     \fi
\ifx \showISSN     \undefined \def \showISSN      #1{\unskip}     \fi
\ifx \showLCCN     \undefined \def \showLCCN      #1{\unskip}     \fi
\ifx \shownote     \undefined \def \shownote      #1{#1}          \fi
\ifx \showarticletitle \undefined \def \showarticletitle #1{#1}   \fi
\ifx \showURL      \undefined \def \showURL       {\relax}        \fi
\providecommand\bibfield[2]{#2}
\providecommand\bibinfo[2]{#2}
\providecommand\natexlab[1]{#1}
\providecommand\showeprint[2][]{arXiv:#2}

\bibitem[Alliance(2022)]%
        {visionserve2022}
\bibfield{author}{\bibinfo{person}{VisionServe Alliance}.} \bibinfo{year}{2022}\natexlab{}.
\newblock \bibinfo{title}{United States' Older Population and Vision Loss: A Briefing}.
\newblock \bibinfo{howpublished}{\url{https://visionservealliance.org}}.
\newblock
\urldef\tempurl%
\url{https://drive.google.com/file/d/1FnyenjMMa4LZNX1gbiaY8klWT-joyZ6D/view}
\showURL{%
\tempurl}
\newblock
\shownote{Prepared by: The Ohio State University, College of Optometry}.


\bibitem[Ault et~al\mbox{.}(2002)]%
        {Ault2002}
\bibfield{author}{\bibinfo{person}{H.~K. Ault}, \bibinfo{person}{J.~W. Deloge}, \bibinfo{person}{R.~W. Lapp}, \bibinfo{person}{M.~J. Morgan}, {and} \bibinfo{person}{J.~R. Barnett}.} \bibinfo{year}{2002}\natexlab{}.
\newblock \showarticletitle{Evaluation of Long Descriptions of Statistical Graphics for Blind and Low Vision Web Users}. In \bibinfo{booktitle}{\emph{Computers Helping People with Special Needs}}, \bibfield{editor}{\bibinfo{person}{Klaus Miesenberger}, \bibinfo{person}{Joachim Klaus}, {and} \bibinfo{person}{Wolfgang Zagler}} (Eds.). \bibinfo{publisher}{Springer Berlin Heidelberg}, \bibinfo{address}{Berlin, Heidelberg}, \bibinfo{pages}{517--526}.
\newblock
\showISBNx{978-3-540-45491-5}


\bibitem[Bru et~al\mbox{.}(2023)]%
        {lineHarp2023}
\bibfield{author}{\bibinfo{person}{Egil Bru}, \bibinfo{person}{Thomas Trautner}, {and} \bibinfo{person}{Stefan Bruckner}.} \bibinfo{year}{2023}\natexlab{}.
\newblock \bibinfo{title}{Line Harp: Importance-Driven Sonification for Dense Line Charts}.
\newblock
\newblock
\showeprint[arxiv]{2307.16589}~[cs.GR]
\urldef\tempurl%
\url{https://arxiv.org/abs/2307.16589}
\showURL{%
\tempurl}


\bibitem[Cherukuru et~al\mbox{.}(2022)]%
        {Nihanth2022}
\bibfield{author}{\bibinfo{person}{Nihanth~W Cherukuru}, \bibinfo{person}{David~A Bailey}, \bibinfo{person}{Tiffany Fourment}, \bibinfo{person}{Becca Hatheway}, \bibinfo{person}{Marika~M Holland}, {and} \bibinfo{person}{Matt Rehme}.} \bibinfo{year}{2022}\natexlab{}.
\newblock \bibinfo{title}{Beyond Visuals : Examining the Experiences of Geoscience Professionals With Vision Disabilities in Accessing Data Visualizations}.
\newblock
\newblock
\showeprint[arxiv]{2207.13220}~[cs.CY]
\urldef\tempurl%
\url{https://arxiv.org/abs/2207.13220}
\showURL{%
\tempurl}


\bibitem[Choi et~al\mbox{.}(9 06)]%
        {Jinho2019}
\bibfield{author}{\bibinfo{person}{Jinho Choi}, \bibinfo{person}{Sanghun Jung}, \bibinfo{person}{Deok~Gun Park}, \bibinfo{person}{Jaegul Choo}, {and} \bibinfo{person}{Niklas Elmqvist}.} \bibinfo{year}{2019-06}\natexlab{}.
\newblock \showarticletitle{Visualizing for the Non‐Visual: Enabling the Visually Impaired to Use Visualization}.
\newblock \bibinfo{journal}{\emph{Computer graphics forum.}} \bibinfo{volume}{38}, \bibinfo{number}{3} (\bibinfo{year}{2019-06}).
\newblock
\showISSN{0167-7055}
\showLCCN{2004233670}


\bibitem[Cleveland and McGill(1984)]%
        {cleveland1984graphical}
\bibfield{author}{\bibinfo{person}{William~S Cleveland} {and} \bibinfo{person}{Robert McGill}.} \bibinfo{year}{1984}\natexlab{}.
\newblock \showarticletitle{Graphical perception: Theory, experimentation, and application to the development of graphical methods}.
\newblock \bibinfo{journal}{\emph{Journal of the American statistical association}} \bibinfo{volume}{79}, \bibinfo{number}{387} (\bibinfo{year}{1984}), \bibinfo{pages}{531--554}.
\newblock


\bibitem[Cohen et~al\mbox{.}(2006)]%
        {Teaching2006}
\bibfield{author}{\bibinfo{person}{Robert~F. Cohen}, \bibinfo{person}{Arthur Meacham}, {and} \bibinfo{person}{Joelle Skaff}.} \bibinfo{year}{2006}\natexlab{}.
\newblock \showarticletitle{Teaching graphs to visually impaired students using an active auditory interface}. In \bibinfo{booktitle}{\emph{Proceedings of the 37th SIGCSE Technical Symposium on Computer Science Education}} (Houston, Texas, USA) \emph{(\bibinfo{series}{SIGCSE '06})}. \bibinfo{publisher}{Association for Computing Machinery}, \bibinfo{address}{New York, NY, USA}, \bibinfo{pages}{279–282}.
\newblock
\showISBNx{1595932593}
\urldef\tempurl%
\url{https://doi.org/10.1145/1121341.1121428}
\showDOI{\tempurl}


\bibitem[Davis et~al\mbox{.}(2022)]%
        {davis2022risks}
\bibfield{author}{\bibinfo{person}{Russell Davis}, \bibinfo{person}{Xiaoying Pu}, \bibinfo{person}{Yiren Ding}, \bibinfo{person}{Brian~D Hall}, \bibinfo{person}{Karen Bonilla}, \bibinfo{person}{Mi Feng}, \bibinfo{person}{Matthew Kay}, {and} \bibinfo{person}{Lane Harrison}.} \bibinfo{year}{2022}\natexlab{}.
\newblock \showarticletitle{The risks of ranking: Revisiting graphical perception to model individual differences in visualization performance}.
\newblock \bibinfo{journal}{\emph{IEEE Transactions on Visualization and Computer Graphics}} \bibinfo{volume}{30}, \bibinfo{number}{3} (\bibinfo{year}{2022}), \bibinfo{pages}{1756--1771}.
\newblock


\bibitem[Demir et~al\mbox{.}(2010)]%
        {Seniz2010}
\bibfield{author}{\bibinfo{person}{Seniz Demir}, \bibinfo{person}{David Oliver}, \bibinfo{person}{Edward Schwartz}, \bibinfo{person}{Stephanie Elzer}, \bibinfo{person}{Sandra Carberry}, \bibinfo{person}{Kathleen~F. Mccoy}, {and} \bibinfo{person}{Daniel Chester}.} \bibinfo{year}{2010}\natexlab{}.
\newblock \showarticletitle{Interactive SIGHT: textual access to simple bar charts}.
\newblock \bibinfo{journal}{\emph{New Review of Hypermedia and Multimedia}} \bibinfo{volume}{16}, \bibinfo{number}{3} (\bibinfo{year}{2010}), \bibinfo{pages}{245--279}.
\newblock
\urldef\tempurl%
\url{https://doi.org/10.1080/13614568.2010.534186}
\showDOI{\tempurl}
\showeprint{https://doi.org/10.1080/13614568.2010.534186}


\bibitem[Elavsky et~al\mbox{.}(2023)]%
        {dataNavigator2023}
\bibfield{author}{\bibinfo{person}{Frank Elavsky}, \bibinfo{person}{Lucas Nadolskis}, {and} \bibinfo{person}{Dominik Moritz}.} \bibinfo{year}{2023}\natexlab{}.
\newblock \bibinfo{title}{Data Navigator: An accessibility-centered data navigation toolkit}.
\newblock
\newblock
\showeprint[arxiv]{2308.08475}~[cs.HC]
\urldef\tempurl%
\url{https://arxiv.org/abs/2308.08475}
\showURL{%
\tempurl}


\bibitem[Engel and Weber(2017a)]%
        {Engel2017}
\bibfield{author}{\bibinfo{person}{Christin Engel} {and} \bibinfo{person}{Gerhard Weber}.} \bibinfo{year}{2017}\natexlab{a}.
\newblock \showarticletitle{Analysis of Tactile Chart Design}. In \bibinfo{booktitle}{\emph{Proceedings of the 10th International Conference on PErvasive Technologies Related to Assistive Environments}} (Island of Rhodes, Greece) \emph{(\bibinfo{series}{PETRA '17})}. \bibinfo{publisher}{Association for Computing Machinery}, \bibinfo{address}{New York, NY, USA}, \bibinfo{pages}{197–200}.
\newblock
\showISBNx{9781450352277}
\urldef\tempurl%
\url{https://doi.org/10.1145/3056540.3064955}
\showDOI{\tempurl}


\bibitem[Engel and Weber(2017b)]%
        {EngelWeber2017}
\bibfield{author}{\bibinfo{person}{Christin Engel} {and} \bibinfo{person}{Gerhard Weber}.} \bibinfo{year}{2017}\natexlab{b}.
\newblock \showarticletitle{Improve the Accessibility of Tactile Charts}. In \bibinfo{booktitle}{\emph{Human-Computer Interaction - INTERACT 2017}}, \bibfield{editor}{\bibinfo{person}{Regina Bernhaupt}, \bibinfo{person}{Girish Dalvi}, \bibinfo{person}{Anirudha Joshi}, \bibinfo{person}{Devanuj K.~Balkrishan}, \bibinfo{person}{Jacki O'Neill}, {and} \bibinfo{person}{Marco Winckler}} (Eds.). \bibinfo{publisher}{Springer International Publishing}, \bibinfo{address}{Cham}, \bibinfo{pages}{187--195}.
\newblock
\showISBNx{978-3-319-67744-6}


\bibitem[Engel and Weber(2018)]%
        {EngelWeber2018}
\bibfield{author}{\bibinfo{person}{Christin Engel} {and} \bibinfo{person}{Gerhard Weber}.} \bibinfo{year}{2018}\natexlab{}.
\newblock \showarticletitle{A User Study to Evaluate Tactile Charts with Blind and Visually Impaired People}. In \bibinfo{booktitle}{\emph{Computers Helping People with Special Needs}}, \bibfield{editor}{\bibinfo{person}{Klaus Miesenberger} {and} \bibinfo{person}{Georgios Kouroupetroglou}} (Eds.). \bibinfo{publisher}{Springer International Publishing}, \bibinfo{address}{Cham}, \bibinfo{pages}{177--184}.
\newblock
\showISBNx{978-3-319-94274-2}


\bibitem[Fan et~al\mbox{.}(2022)]%
        {Danyang2022}
\bibfield{author}{\bibinfo{person}{Danyang Fan}, \bibinfo{person}{Alexa~Fay Siu}, \bibinfo{person}{Wing-Sum~Adrienne Law}, \bibinfo{person}{Raymond~Ruihong Zhen}, \bibinfo{person}{Sile O'Modhrain}, {and} \bibinfo{person}{Sean Follmer}.} \bibinfo{year}{2022}\natexlab{}.
\newblock \showarticletitle{Slide-Tone and Tilt-Tone: 1-DOF Haptic Techniques for Conveying Shape Characteristics of Graphs to Blind Users}. In \bibinfo{booktitle}{\emph{Proceedings of the 2022 CHI Conference on Human Factors in Computing Systems}} (New Orleans, LA, USA) \emph{(\bibinfo{series}{CHI '22})}. \bibinfo{publisher}{Association for Computing Machinery}, \bibinfo{address}{New York, NY, USA}, Article \bibinfo{articleno}{477}, \bibinfo{numpages}{19}~pages.
\newblock
\showISBNx{9781450391573}
\urldef\tempurl%
\url{https://doi.org/10.1145/3491102.3517790}
\showDOI{\tempurl}


\bibitem[Fritz and Barner(1999)]%
        {Fritz1999}
\bibfield{author}{\bibinfo{person}{J.P. Fritz} {and} \bibinfo{person}{K.E. Barner}.} \bibinfo{year}{1999}\natexlab{}.
\newblock \showarticletitle{Design of a haptic data visualization system for people with visual impairments}.
\newblock \bibinfo{journal}{\emph{IEEE Transactions on Rehabilitation Engineering}} \bibinfo{volume}{7}, \bibinfo{number}{3} (\bibinfo{year}{1999}), \bibinfo{pages}{372--384}.
\newblock
\urldef\tempurl%
\url{https://doi.org/10.1109/86.788473}
\showDOI{\tempurl}


\bibitem[Gelman et~al\mbox{.}(2020)]%
        {gelman2020bayesian}
\bibfield{author}{\bibinfo{person}{Andrew Gelman}, \bibinfo{person}{Aki Vehtari}, \bibinfo{person}{Daniel Simpson}, \bibinfo{person}{Charles~C Margossian}, \bibinfo{person}{Bob Carpenter}, \bibinfo{person}{Yuling Yao}, \bibinfo{person}{Lauren Kennedy}, \bibinfo{person}{Jonah Gabry}, \bibinfo{person}{Paul-Christian B{\"u}rkner}, {and} \bibinfo{person}{Martin Modr{\'a}k}.} \bibinfo{year}{2020}\natexlab{}.
\newblock \showarticletitle{Bayesian workflow}.
\newblock \bibinfo{journal}{\emph{arXiv preprint arXiv:2011.01808}} (\bibinfo{year}{2020}).
\newblock


\bibitem[Goncu et~al\mbox{.}(2010)]%
        {Goncu2010}
\bibfield{author}{\bibinfo{person}{Cagatay Goncu}, \bibinfo{person}{Kim Marriott}, {and} \bibinfo{person}{John Hurst}.} \bibinfo{year}{2010}\natexlab{}.
\newblock \showarticletitle{Usability of Accessible Bar Charts}. In \bibinfo{booktitle}{\emph{Diagrammatic Representation and Inference}}, \bibfield{editor}{\bibinfo{person}{Ashok~K. Goel}, \bibinfo{person}{Mateja Jamnik}, {and} \bibinfo{person}{N.~Hari Narayanan}} (Eds.). \bibinfo{publisher}{Springer Berlin Heidelberg}, \bibinfo{address}{Berlin, Heidelberg}, \bibinfo{pages}{167--181}.
\newblock
\showISBNx{978-3-642-14600-8}


\bibitem[Guinness et~al\mbox{.}(2019)]%
        {Darren2019}
\bibfield{author}{\bibinfo{person}{Darren Guinness}, \bibinfo{person}{Annika Muehlbradt}, \bibinfo{person}{Daniel Szafir}, {and} \bibinfo{person}{Shaun~K. Kane}.} \bibinfo{year}{2019}\natexlab{}.
\newblock \showarticletitle{RoboGraphics: Dynamic Tactile Graphics Powered by Mobile Robots}. In \bibinfo{booktitle}{\emph{Proceedings of the 21st International ACM SIGACCESS Conference on Computers and Accessibility}} (Pittsburgh, PA, USA) \emph{(\bibinfo{series}{ASSETS '19})}. \bibinfo{publisher}{Association for Computing Machinery}, \bibinfo{address}{New York, NY, USA}, \bibinfo{pages}{318–328}.
\newblock
\showISBNx{9781450366762}
\urldef\tempurl%
\url{https://doi.org/10.1145/3308561.3353804}
\showDOI{\tempurl}


\bibitem[Heer and Bostock(2010)]%
        {heer2010crowdsourcing}
\bibfield{author}{\bibinfo{person}{Jeffrey Heer} {and} \bibinfo{person}{Michael Bostock}.} \bibinfo{year}{2010}\natexlab{}.
\newblock \showarticletitle{Crowdsourcing graphical perception: using mechanical turk to assess visualization design}. In \bibinfo{booktitle}{\emph{Proceedings of the SIGCHI conference on human factors in computing systems}}. \bibinfo{pages}{203--212}.
\newblock


\bibitem[Kildal and Brewster(2006)]%
        {Non-visual2006}
\bibfield{author}{\bibinfo{person}{Johan Kildal} {and} \bibinfo{person}{Stephen~A. Brewster}.} \bibinfo{year}{2006}\natexlab{}.
\newblock \showarticletitle{Non-visual overviews of complex data sets}. In \bibinfo{booktitle}{\emph{CHI '06 Extended Abstracts on Human Factors in Computing Systems}} (Montr\'{e}al, Qu\'{e}bec, Canada) \emph{(\bibinfo{series}{CHI EA '06})}. \bibinfo{publisher}{Association for Computing Machinery}, \bibinfo{address}{New York, NY, USA}, \bibinfo{pages}{947–952}.
\newblock
\showISBNx{1595932984}
\urldef\tempurl%
\url{https://doi.org/10.1145/1125451.1125634}
\showDOI{\tempurl}


\bibitem[Kim and Lim(2011a)]%
        {Dajungkim2011}
\bibfield{author}{\bibinfo{person}{Da-jung Kim} {and} \bibinfo{person}{Youn-kyung Lim}.} \bibinfo{year}{2011}\natexlab{a}.
\newblock \showarticletitle{Handscope: enabling blind people to experience statistical graphics on websites through haptics}. In \bibinfo{booktitle}{\emph{Proceedings of the SIGCHI Conference on Human Factors in Computing Systems}} (Vancouver, BC, Canada) \emph{(\bibinfo{series}{CHI '11})}. \bibinfo{publisher}{Association for Computing Machinery}, \bibinfo{address}{New York, NY, USA}, \bibinfo{pages}{2039–2042}.
\newblock
\showISBNx{9781450302289}
\urldef\tempurl%
\url{https://doi.org/10.1145/1978942.1979237}
\showDOI{\tempurl}


\bibitem[Kim and Lim(2011b)]%
        {dajung20211}
\bibfield{author}{\bibinfo{person}{Da-jung Kim} {and} \bibinfo{person}{Youn-kyung Lim}.} \bibinfo{year}{2011}\natexlab{b}.
\newblock \showarticletitle{Handscope: enabling blind people to experience statistical graphics on websites through haptics}. In \bibinfo{booktitle}{\emph{Proceedings of the SIGCHI Conference on Human Factors in Computing Systems}} (Vancouver, BC, Canada) \emph{(\bibinfo{series}{CHI '11})}. \bibinfo{publisher}{Association for Computing Machinery}, \bibinfo{address}{New York, NY, USA}, \bibinfo{pages}{2039–2042}.
\newblock
\showISBNx{9781450302289}
\urldef\tempurl%
\url{https://doi.org/10.1145/1978942.1979237}
\showDOI{\tempurl}


\bibitem[Kim et~al\mbox{.}(2024)]%
        {erie2024}
\bibfield{author}{\bibinfo{person}{Hyeok Kim}, \bibinfo{person}{Yea-Seul Kim}, {and} \bibinfo{person}{Jessica Hullman}.} \bibinfo{year}{2024}\natexlab{}.
\newblock \showarticletitle{Erie: A Declarative Grammar for Data Sonification}. In \bibinfo{booktitle}{\emph{Proceedings of the CHI Conference on Human Factors in Computing Systems}} (Honolulu, HI, USA) \emph{(\bibinfo{series}{CHI '24})}. \bibinfo{publisher}{Association for Computing Machinery}, \bibinfo{address}{New York, NY, USA}, Article \bibinfo{articleno}{986}, \bibinfo{numpages}{19}~pages.
\newblock
\showISBNx{9798400703300}
\urldef\tempurl%
\url{https://doi.org/10.1145/3613904.3642442}
\showDOI{\tempurl}


\bibitem[Kim et~al\mbox{.}(2023)]%
        {kim2023exploring}
\bibfield{author}{\bibinfo{person}{Jiho Kim}, \bibinfo{person}{Arjun Srinivasan}, \bibinfo{person}{Nam~Wook Kim}, {and} \bibinfo{person}{Yea-Seul Kim}.} \bibinfo{year}{2023}\natexlab{}.
\newblock \showarticletitle{Exploring chart question answering for blind and low vision users}. In \bibinfo{booktitle}{\emph{Proceedings of the 2023 CHI Conference on Human Factors in Computing Systems}}. \bibinfo{pages}{1--15}.
\newblock


\bibitem[Ladner et~al\mbox{.}(2005)]%
        {Automating2005}
\bibfield{author}{\bibinfo{person}{Richard~E. Ladner}, \bibinfo{person}{Melody~Y. Ivory}, \bibinfo{person}{Rajesh Rao}, \bibinfo{person}{Sheryl Burgstahler}, \bibinfo{person}{Dan Comden}, \bibinfo{person}{Sangyun Hahn}, \bibinfo{person}{Matthew Renzelmann}, \bibinfo{person}{Satria Krisnandi}, \bibinfo{person}{Mahalakshmi Ramasamy}, \bibinfo{person}{Beverly Slabosky}, \bibinfo{person}{Andrew Martin}, \bibinfo{person}{Amelia Lacenski}, \bibinfo{person}{Stuart Olsen}, {and} \bibinfo{person}{Dmitri Groce}.} \bibinfo{year}{2005}\natexlab{}.
\newblock \showarticletitle{Automating tactile graphics translation}. In \bibinfo{booktitle}{\emph{Proceedings of the 7th International ACM SIGACCESS Conference on Computers and Accessibility}} (Baltimore, MD, USA) \emph{(\bibinfo{series}{Assets '05})}. \bibinfo{publisher}{Association for Computing Machinery}, \bibinfo{address}{New York, NY, USA}, \bibinfo{pages}{150–157}.
\newblock
\showISBNx{1595931597}
\urldef\tempurl%
\url{https://doi.org/10.1145/1090785.1090814}
\showDOI{\tempurl}


\bibitem[Lundgard et~al\mbox{.}(2019)]%
        {Alan2019}
\bibfield{author}{\bibinfo{person}{Alan Lundgard}, \bibinfo{person}{Crystal Lee}, {and} \bibinfo{person}{Arvind Satyanarayan}.} \bibinfo{year}{2019}\natexlab{}.
\newblock \showarticletitle{Sociotechnical Considerations for Accessible Visualization Design}. In \bibinfo{booktitle}{\emph{2019 IEEE Visualization Conference (VIS)}}. \bibinfo{pages}{16--20}.
\newblock
\urldef\tempurl%
\url{https://doi.org/10.1109/VISUAL.2019.8933762}
\showDOI{\tempurl}


\bibitem[Lundgard and Satyanarayan(2021)]%
        {lundgard2021accessible}
\bibfield{author}{\bibinfo{person}{Alan Lundgard} {and} \bibinfo{person}{Arvind Satyanarayan}.} \bibinfo{year}{2021}\natexlab{}.
\newblock \showarticletitle{Accessible visualization via natural language descriptions: A four-level model of semantic content}.
\newblock \bibinfo{journal}{\emph{IEEE transactions on visualization and computer graphics}} \bibinfo{volume}{28}, \bibinfo{number}{1} (\bibinfo{year}{2021}), \bibinfo{pages}{1073--1083}.
\newblock


\bibitem[Marriott et~al\mbox{.}(2021)]%
        {Marriott2021}
\bibfield{author}{\bibinfo{person}{Kim Marriott}, \bibinfo{person}{Bongshin Lee}, \bibinfo{person}{Matthew Butler}, \bibinfo{person}{Ed Cutrell}, \bibinfo{person}{Kirsten Ellis}, \bibinfo{person}{Cagatay Goncu}, \bibinfo{person}{Marti Hearst}, \bibinfo{person}{Kathleen McCoy}, {and} \bibinfo{person}{Danielle~Albers Szafir}.} \bibinfo{year}{2021}\natexlab{}.
\newblock \showarticletitle{Inclusive data visualization for people with disabilities: a call to action}.
\newblock \bibinfo{journal}{\emph{Interactions}} \bibinfo{volume}{28}, \bibinfo{number}{3} (\bibinfo{date}{apr} \bibinfo{year}{2021}), \bibinfo{pages}{47–51}.
\newblock
\showISSN{1072-5520}
\urldef\tempurl%
\url{https://doi.org/10.1145/3457875}
\showDOI{\tempurl}


\bibitem[McDonnall and Sui(2019)]%
        {mcdonnall2019employment}
\bibfield{author}{\bibinfo{person}{Michele~C. McDonnall} {and} \bibinfo{person}{Zhen Sui}.} \bibinfo{year}{2019}\natexlab{}.
\newblock \showarticletitle{Employment and Unemployment Rates of People who are Blind or Visually Impaired}.
\newblock \bibinfo{journal}{\emph{Journal of Visual Impairment \& Blindness}} \bibinfo{volume}{113}, \bibinfo{number}{3} (\bibinfo{year}{2019}), \bibinfo{pages}{245--254}.
\newblock


\bibitem[Mishra et~al\mbox{.}(2022)]%
        {ChartVi2022}
\bibfield{author}{\bibinfo{person}{Prerna Mishra}, \bibinfo{person}{Santosh Kumar}, \bibinfo{person}{Mithilesh~Kumar Chaube}, {and} \bibinfo{person}{Urmila Shrawankar}.} \bibinfo{year}{2022}\natexlab{}.
\newblock \showarticletitle{ChartVi: Charts summarizer for visually impaired}.
\newblock \bibinfo{journal}{\emph{Journal of Computer Languages}}  \bibinfo{volume}{69} (\bibinfo{year}{2022}), \bibinfo{pages}{101107}.
\newblock
\showISSN{2590-1184}
\urldef\tempurl%
\url{https://doi.org/10.1016/j.cola.2022.101107}
\showDOI{\tempurl}


\bibitem[Reinders et~al\mbox{.}(2024)]%
        {reinders2024refreshable}
\bibfield{author}{\bibinfo{person}{Samuel Reinders}, \bibinfo{person}{Matthew Butler}, \bibinfo{person}{Ingrid Zukerman}, \bibinfo{person}{Bongshin Lee}, \bibinfo{person}{Lizhen Qu}, {and} \bibinfo{person}{Kim Marriott}.} \bibinfo{year}{2024}\natexlab{}.
\newblock \showarticletitle{When Refreshable Tactile Displays Meet Conversational Agents: Investigating Accessible Data Presentation and Analysis with Touch and Speech}.
\newblock \bibinfo{journal}{\emph{arXiv preprint arXiv:2408.04806}} (\bibinfo{year}{2024}).
\newblock


\bibitem[Rosenberg(2021)]%
        {lighthouseFlattenCurve2021}
\bibfield{author}{\bibinfo{person}{Naomi Rosenberg}.} \bibinfo{year}{2021}\natexlab{}.
\newblock \bibinfo{booktitle}{\emph{What does Flattening the Curve look like?}}
\newblock
\urldef\tempurl%
\url{https://lighthouse-sf.org/2021/04/30/flattening-the-curve/}
\showURL{%
\tempurl}


\bibitem[Rowell and Ungar(2003)]%
        {Jonathan2003}
\bibfield{author}{\bibinfo{person}{Jonathan Rowell} {and} \bibinfo{person}{Simon Ungar}.} \bibinfo{year}{2003}\natexlab{}.
\newblock \showarticletitle{The world of touch: an international survey of tactile maps. Part 1: production}.
\newblock \bibinfo{journal}{\emph{British Journal of Visual Impairment}} \bibinfo{volume}{21}, \bibinfo{number}{3} (\bibinfo{year}{2003}), \bibinfo{pages}{98--104}.
\newblock
\urldef\tempurl%
\url{https://doi.org/10.1177/026461960302100303}
\showDOI{\tempurl}
\showeprint{https://doi.org/10.1177/026461960302100303}


\bibitem[Rowell and Ungar(2005)]%
        {rowell2005feeling}
\bibfield{author}{\bibinfo{person}{Jonathan Rowell} {and} \bibinfo{person}{Simon Ungar}.} \bibinfo{year}{2005}\natexlab{}.
\newblock \showarticletitle{Feeling our way: tactile map user requirements-a survey}. In \bibinfo{booktitle}{\emph{International Cartographic Conference, La Coruna}}, Vol.~\bibinfo{volume}{152}.
\newblock


\bibitem[Sharif et~al\mbox{.}(2021)]%
        {Ather2021}
\bibfield{author}{\bibinfo{person}{Ather Sharif}, \bibinfo{person}{Sanjana~Shivani Chintalapati}, \bibinfo{person}{Jacob~O. Wobbrock}, {and} \bibinfo{person}{Katharina Reinecke}.} \bibinfo{year}{2021}\natexlab{}.
\newblock \showarticletitle{Understanding Screen-Reader Users’ Experiences with Online Data Visualizations}. In \bibinfo{booktitle}{\emph{Proceedings of the 23rd International ACM SIGACCESS Conference on Computers and Accessibility}} (Virtual Event, USA) \emph{(\bibinfo{series}{ASSETS '21})}. \bibinfo{publisher}{Association for Computing Machinery}, \bibinfo{address}{New York, NY, USA}, Article \bibinfo{articleno}{14}, \bibinfo{numpages}{16}~pages.
\newblock
\showISBNx{9781450383066}
\urldef\tempurl%
\url{https://doi.org/10.1145/3441852.3471202}
\showDOI{\tempurl}


\bibitem[Sharif et~al\mbox{.}(2022)]%
        {Ather2022}
\bibfield{author}{\bibinfo{person}{Ather Sharif}, \bibinfo{person}{Olivia~H. Wang}, \bibinfo{person}{Alida~T. Muongchan}, \bibinfo{person}{Katharina Reinecke}, {and} \bibinfo{person}{Jacob~O. Wobbrock}.} \bibinfo{year}{2022}\natexlab{}.
\newblock \showarticletitle{VoxLens: Making Online Data Visualizations Accessible with an Interactive JavaScript Plug-In}. In \bibinfo{booktitle}{\emph{Proceedings of the 2022 CHI Conference on Human Factors in Computing Systems}} (New Orleans, LA, USA) \emph{(\bibinfo{series}{CHI '22})}. \bibinfo{publisher}{Association for Computing Machinery}, \bibinfo{address}{New York, NY, USA}, Article \bibinfo{articleno}{478}, \bibinfo{numpages}{19}~pages.
\newblock
\showISBNx{9781450391573}
\urldef\tempurl%
\url{https://doi.org/10.1145/3491102.3517431}
\showDOI{\tempurl}


\bibitem[Shneiderman(2003)]%
        {shneiderman2003eyes}
\bibfield{author}{\bibinfo{person}{Ben Shneiderman}.} \bibinfo{year}{2003}\natexlab{}.
\newblock \showarticletitle{The eyes have it: A task by data type taxonomy for information visualizations}.
\newblock In \bibinfo{booktitle}{\emph{The craft of information visualization}}. \bibinfo{publisher}{Elsevier}, \bibinfo{pages}{364--371}.
\newblock


\bibitem[Siu et~al\mbox{.}(2022)]%
        {Alexa2022}
\bibfield{author}{\bibinfo{person}{Alexa Siu}, \bibinfo{person}{Gene S-H~Kim}, \bibinfo{person}{Sile O'Modhrain}, {and} \bibinfo{person}{Sean Follmer}.} \bibinfo{year}{2022}\natexlab{}.
\newblock \showarticletitle{Supporting Accessible Data Visualization Through Audio Data Narratives}. In \bibinfo{booktitle}{\emph{Proceedings of the 2022 CHI Conference on Human Factors in Computing Systems}} (New Orleans, LA, USA) \emph{(\bibinfo{series}{CHI '22})}. \bibinfo{publisher}{Association for Computing Machinery}, \bibinfo{address}{New York, NY, USA}, Article \bibinfo{articleno}{476}, \bibinfo{numpages}{19}~pages.
\newblock
\showISBNx{9781450391573}
\urldef\tempurl%
\url{https://doi.org/10.1145/3491102.3517678}
\showDOI{\tempurl}


\bibitem[Tabrik(2022)]%
        {tabrik2022}
\bibfield{author}{\bibinfo{person}{S. Tabrik}.} \bibinfo{year}{2022}\natexlab{}.
\newblock \showarticletitle{Neural Mechanisms Underlying Cross-Modal Object Categorization: Visual and Tactile Senses}.
\newblock  (\bibinfo{year}{2022}).
\newblock
\urldef\tempurl%
\url{https://scholar.archive.org/work/i6f56afzgba6xbxadt6rztwita/access/wayback/https://hss-opus.ub.ruhr-uni-bochum.de/opus4/frontdoor/deliver/index/docId/9089/file/diss.pdf}
\showURL{%
\tempurl}


\bibitem[Talbot et~al\mbox{.}(2014)]%
        {Talbot2014}
\bibfield{author}{\bibinfo{person}{Justin Talbot}, \bibinfo{person}{Vidya Setlur}, {and} \bibinfo{person}{Anushka Anand}.} \bibinfo{year}{2014}\natexlab{}.
\newblock \showarticletitle{Four Experiments on the Perception of Bar Charts}.
\newblock \bibinfo{journal}{\emph{IEEE Transactions on Visualization and Computer Graphics}} \bibinfo{volume}{20}, \bibinfo{number}{12} (\bibinfo{year}{2014}), \bibinfo{pages}{2152--2160}.
\newblock
\urldef\tempurl%
\url{https://doi.org/10.1109/TVCG.2014.2346320}
\showDOI{\tempurl}


\bibitem[Thompson et~al\mbox{.}(2023)]%
        {chartreader}
\bibfield{author}{\bibinfo{person}{John~R Thompson}, \bibinfo{person}{Jesse~J Martinez}, \bibinfo{person}{Alper Sarikaya}, \bibinfo{person}{Edward Cutrell}, {and} \bibinfo{person}{Bongshin Lee}.} \bibinfo{year}{2023}\natexlab{}.
\newblock \showarticletitle{Chart Reader: Accessible Visualization Experiences Designed with Screen Reader Users}. In \bibinfo{booktitle}{\emph{Proceedings of the 2023 CHI Conference on Human Factors in Computing Systems}} (Hamburg, Germany) \emph{(\bibinfo{series}{CHI '23})}. \bibinfo{publisher}{Association for Computing Machinery}, \bibinfo{address}{New York, NY, USA}, Article \bibinfo{articleno}{802}, \bibinfo{numpages}{18}~pages.
\newblock
\showISBNx{9781450394215}
\urldef\tempurl%
\url{https://doi.org/10.1145/3544548.3581186}
\showDOI{\tempurl}


\bibitem[Thunstr{\"o}m et~al\mbox{.}(2020)]%
        {thunstrom2020benefits}
\bibfield{author}{\bibinfo{person}{Linda Thunstr{\"o}m}, \bibinfo{person}{Stephen~C Newbold}, \bibinfo{person}{David Finnoff}, \bibinfo{person}{Madison Ashworth}, {and} \bibinfo{person}{Jason~F Shogren}.} \bibinfo{year}{2020}\natexlab{}.
\newblock \showarticletitle{The benefits and costs of using social distancing to flatten the curve for COVID-19}.
\newblock \bibinfo{journal}{\emph{Journal of Benefit-Cost Analysis}} \bibinfo{volume}{11}, \bibinfo{number}{2} (\bibinfo{year}{2020}), \bibinfo{pages}{179--195}.
\newblock


\bibitem[Tufte and Graves-Morris(1983)]%
        {tufte1983visual}
\bibfield{author}{\bibinfo{person}{Edward~R Tufte} {and} \bibinfo{person}{Peter~R Graves-Morris}.} \bibinfo{year}{1983}\natexlab{}.
\newblock \bibinfo{booktitle}{\emph{The visual display of quantitative information}}. Vol.~\bibinfo{volume}{2}.
\newblock \bibinfo{publisher}{Graphics press Cheshire, CT}.
\newblock


\bibitem[Wall and Brewster(2006)]%
        {steven2006}
\bibfield{author}{\bibinfo{person}{Steven Wall} {and} \bibinfo{person}{Stephen Brewster}.} \bibinfo{year}{2006}\natexlab{}.
\newblock \showarticletitle{Feeling what you hear: tactile feedback for navigation of audio graphs}. In \bibinfo{booktitle}{\emph{Proceedings of the SIGCHI Conference on Human Factors in Computing Systems}} (Montr\'{e}al, Qu\'{e}bec, Canada) \emph{(\bibinfo{series}{CHI '06})}. \bibinfo{publisher}{Association for Computing Machinery}, \bibinfo{address}{New York, NY, USA}, \bibinfo{pages}{1123–1132}.
\newblock
\showISBNx{1595933727}
\urldef\tempurl%
\url{https://doi.org/10.1145/1124772.1124941}
\showDOI{\tempurl}


\bibitem[Wang et~al\mbox{.}(2 06)]%
        {WangR2022}
\bibfield{author}{\bibinfo{person}{R Wang}, \bibinfo{person}{C Jung}, {and} \bibinfo{person}{Y Kim}.} \bibinfo{year}{2022-06}\natexlab{}.
\newblock \showarticletitle{Seeing Through Sounds: Mapping Auditory Dimensions to Data and Charts for People with Visual Impairments}.
\newblock \bibinfo{journal}{\emph{Computer graphics forum.}} \bibinfo{volume}{41}, \bibinfo{number}{3} (\bibinfo{year}{2022-06}).
\newblock
\showISSN{0167-7055}
\showLCCN{2004233670}


\bibitem[Watanabe and Inaba(2018)]%
        {Watanabe2018}
\bibfield{author}{\bibinfo{person}{Tetsuya Watanabe} {and} \bibinfo{person}{Naoki Inaba}.} \bibinfo{year}{2018}\natexlab{}.
\newblock \showarticletitle{Textures Suitable for Tactile Bar Charts on Capsule Paper}.
\newblock \bibinfo{journal}{\emph{Transactions of the Virtual Reality Society of Japan}} \bibinfo{volume}{23}, \bibinfo{number}{1} (\bibinfo{year}{2018}), \bibinfo{pages}{13--20}.
\newblock
\urldef\tempurl%
\url{https://doi.org/10.18974/tvrsj.23.1_13}
\showDOI{\tempurl}


\bibitem[Watanabe and Mizukami(2018)]%
        {WatanabeMizukami2018}
\bibfield{author}{\bibinfo{person}{Tetsuya Watanabe} {and} \bibinfo{person}{Hikaru Mizukami}.} \bibinfo{year}{2018}\natexlab{}.
\newblock \showarticletitle{Effectiveness of Tactile Scatter Plots: Comparison of Non-visual Data Representations}. In \bibinfo{booktitle}{\emph{Computers Helping People with Special Needs}}, \bibfield{editor}{\bibinfo{person}{Klaus Miesenberger} {and} \bibinfo{person}{Georgios Kouroupetroglou}} (Eds.). \bibinfo{publisher}{Springer International Publishing}, \bibinfo{address}{Cham}, \bibinfo{pages}{628--635}.
\newblock
\showISBNx{978-3-319-94277-3}


\bibitem[Wong(2010)]%
        {wong2010points}
\bibfield{author}{\bibinfo{person}{Bang Wong}.} \bibinfo{year}{2010}\natexlab{}.
\newblock \showarticletitle{Points of view: Gestalt principles (Part 1)}.
\newblock \bibinfo{journal}{\emph{nature methods}} \bibinfo{volume}{7}, \bibinfo{number}{11} (\bibinfo{year}{2010}), \bibinfo{pages}{863}.
\newblock


\bibitem[Yang et~al\mbox{.}(2020)]%
        {Yalong2020}
\bibfield{author}{\bibinfo{person}{Yalong Yang}, \bibinfo{person}{Kim Marriott}, \bibinfo{person}{Matthew Butler}, \bibinfo{person}{Cagatay Goncu}, {and} \bibinfo{person}{Leona Holloway}.} \bibinfo{year}{2020}\natexlab{}.
\newblock \showarticletitle{Tactile Presentation of Network Data: Text, Matrix or Diagram?}. In \bibinfo{booktitle}{\emph{Proceedings of the 2020 CHI Conference on Human Factors in Computing Systems}} (Honolulu, HI, USA) \emph{(\bibinfo{series}{CHI '20})}. \bibinfo{publisher}{Association for Computing Machinery}, \bibinfo{address}{New York, NY, USA}, \bibinfo{pages}{1–12}.
\newblock
\showISBNx{9781450367080}
\urldef\tempurl%
\url{https://doi.org/10.1145/3313831.3376367}
\showDOI{\tempurl}


\bibitem[Yu et~al\mbox{.}(2001)]%
        {Haptic2001}
\bibfield{author}{\bibinfo{person}{Wai Yu}, \bibinfo{person}{Ramesh Ramloll}, {and} \bibinfo{person}{Stephen Brewster}.} \bibinfo{year}{2001}\natexlab{}.
\newblock \showarticletitle{Haptic graphs for blind computer users}. In \bibinfo{booktitle}{\emph{Haptic Human-Computer Interaction}}, \bibfield{editor}{\bibinfo{person}{Stephen Brewster} {and} \bibinfo{person}{Roderick Murray-Smith}} (Eds.). \bibinfo{publisher}{Springer Berlin Heidelberg}, \bibinfo{address}{Berlin, Heidelberg}, \bibinfo{pages}{41--51}.
\newblock
\showISBNx{978-3-540-44589-0}


\bibitem[Zhao et~al\mbox{.}(2008)]%
        {Haixia2008}
\bibfield{author}{\bibinfo{person}{Haixia Zhao}, \bibinfo{person}{Catherine Plaisant}, \bibinfo{person}{Ben Shneiderman}, {and} \bibinfo{person}{Jonathan Lazar}.} \bibinfo{year}{2008}\natexlab{}.
\newblock \showarticletitle{Data Sonification for Users with Visual Impairment: A Case Study with Georeferenced Data}.
\newblock \bibinfo{journal}{\emph{ACM Trans. Comput.-Hum. Interact.}} \bibinfo{volume}{15}, \bibinfo{number}{1}, Article \bibinfo{articleno}{4} (\bibinfo{date}{may} \bibinfo{year}{2008}), \bibinfo{numpages}{28}~pages.
\newblock
\showISSN{1073-0516}
\urldef\tempurl%
\url{https://doi.org/10.1145/1352782.1352786}
\showDOI{\tempurl}


\bibitem[Zong et~al\mbox{.}(2022)]%
        {zong2022rich}
\bibfield{author}{\bibinfo{person}{Jonathan Zong}, \bibinfo{person}{Crystal Lee}, \bibinfo{person}{Alan Lundgard}, \bibinfo{person}{JiWoong Jang}, \bibinfo{person}{Daniel Hajas}, {and} \bibinfo{person}{Arvind Satyanarayan}.} \bibinfo{year}{2022}\natexlab{}.
\newblock \showarticletitle{Rich screen reader experiences for accessible data visualization}. In \bibinfo{booktitle}{\emph{Computer Graphics Forum}}, Vol.~\bibinfo{volume}{41}. Wiley Online Library, \bibinfo{pages}{15--27}.
\newblock


\end{thebibliography}

\appendix

\end{document}